\documentclass[usenatbib,times,usegraphicx,useAMS]{mn2e}
\usepackage{amssymb,url,times,bm}

{\left\lbrace\begin{array}{@{}l@{}}}%
{\end{array}\right.}

\defcitealias{facchini13_1}{FLP}
\defcitealias{foucart_lai_13}{Foucart \& Lai 2013}

\title[Probing planets in transition discs' cavities via warps]{Probing the presence of planets in transition discs' cavities via warps: the case of TW Hya}

\author[Facchini, Ricci and Lodato]{Stefano Facchini$^{1}$\thanks{facchini@ast.cam.ac.uk}, Luca Ricci$^2$ and Giuseppe Lodato$^{3}$\\
$^1$Institute of Astronomy, Madingley Road, Cambridge CB3 OHA, UK\\
$^2$California Institute of Technology, 1200 East California Boulervard, 91125 Pasadena, CA, USA\\
$^3$Dipartimento di Fisica, Universit\`a Degli Studi di Milano, Via Celoria, 16, Milano, 20133, Italy
}
\date{Submission date}
\pagerange{\pageref{firstpage}--\pageref{lastpage}} \pubyear{2014}

\begin{document}
\label{firstpage}
\bibliographystyle{mn2e}
\maketitle

\begin{abstract}

We are entering the era in which observations of protoplanetary discs properties can indirectly probe the presence of massive planets or low mass stellar companions interacting with the disc. In particular, the detection of warped discs can provide important clues to the properties of the star-disc system. In this paper we show how observations of warped discs can be used to infer the dynamical properties of the systems. We concentrate on circumbinary discs, where the mass of the secondary can be planetary. First, we provide some simple relations that link the amplitude of the warp in the linear regime to the parameters of the system. Secondly, we apply our method to the case of TW Hya, a transition disc for which a warp has been proposed based on spectroscopic observations. Assuming values for the disc and stellar parameters from observations, we conclude that, in order for a warp induced by a planetary companion to be detectable, the planet mass should be large ($M_{\rm p}\approx 10-14 M_{\rm J}$) and the disc should be viscous ($\alpha\approx 0.15-0.25$). We also apply our model to LkCa 15 and T Cha, where a substellar companion has been detected within the central cavity of the transition discs.

\end{abstract}



\section{Introduction}
\label{sec:introduction}

It is now well known that planets form in protostellar discs \citep[e.g.][for a review]{armitage_11}. However, it is still observationally challenging to observe planets in the act of formation, i.e. when they are still embedded or surrounded by a protoplanetary disc. On the other hand, it is possible to indirectly probe (or at least infer) the presence of a planet by looking at the dynamical effects that it causes onto the disc. As an example, \citet{garufi_13} have recently suggested the presence of a planet in HD135344B by analysing the spiral structure and radial distribution of different grain sizes in the disc. Another effect that planets can have on discs are azimuthal asymmetries in the gas and dust profiles \citep[e.g.][on the discs surrounding the young stars LkH$\alpha$ 330 and Oph IRS 48, respectively]{isella_13,van_der_marel_13}.

These examples are all referred to transition discs, i.e. discs that present either a gap or a cleared central cavity \citep[see review by][]{espaillat_14}. This is due to two main reasons. A first one is that these discs are thought to be at the end of their lifetime, therefore they are more likely to have already formed planets. The second reason is that transition discs are thought to be a transient phase of the disc life, just before the final dispersal. They are therefore unique candidates where to probe the physical mechanism originating the dispersal. Many models have been invoked to explain such a transition phase \citep[see the review by][]{alexander_13}. Among these models we can mention photoevaporation, either internally \citep[e.g.][]{owen_12} or externally driven \citep[e.g.][]{adams_04}, grain growth \citep[e.g.][]{birnstiel_12} and magnetic winds \citep[e.g.][]{armitage_13}. Another leading model is that the central cavity of transition discs is cleared out by dynamical interactions with one or more planets \citep[e.g.][]{zhu_11}. Being able to observe them either directly or indirectly in the central cavity would be fundamental to discriminate between these different scenarios. Note that some candidate protoplanets have already been directly observed in the cavity of transition discs \citep[e.g.][in T Cha and LkCa 15, respectively]{huelamo_11,kraus_12}.

One dynamical effect that has been theoretically addressed recently is disc warping due to the presence of misaligned planets \citep[e.g.][]{xiang_13,bitsch_13}, i.e. planets orbiting on a plane that is misaligned with respect to the outer regions of the disc. Misaligned planets seem likely to occur. A few resonant mechanisms are known to be responsible for this effect. For example, planet-planet scattering, in the case of multi-planetary systems, appears to be effective even when planets are still embedded in the disc \citep[][]{lega_13}. Secondly, resonant interactions during migration can excite planets outside the coplanar orbit. Finally, a misaligned planet could simply be tracing a past inclined structure of the disc itself \citep[e.g.][]{bate10}. Mutual inclinations of planets with respect to the disc plane can then be maintained  by the Kozai mechanism  \citep[e.g.][]{teyssandier_13}. The dynamical response of the disc to the secular torque produced by misaligned (proto)planets (and vice-versa) has been analysed mainly by numerical simulations for the case in which the planet is embedded in the disc \citep[though][used an analytical approach]{terquem_13}. Note that we refer to `embedded' planet as opposed to a planet that is in the central cavity of the disc. \citet{facchini13_1} have developed a simple 1D model obtaining the steady-state solution for the warped structure of circumbinary discs. The model was developed to deal with stellar binaries, but it is applicable to planetary systems where a massive planet is orbiting a star within a central cavity of the disc.

Despite the theory has been quite developed, only a few cases of warped structures have been observed either in  gas-rich or in gas-poor discs, since measuring the warping of a disc is very challenging. The famous case of  $\beta$-Pic lies in this small sample. Many images have shown a warped structure in this debris disc \citep[e.g.][]{heap_00,golimowski_06}. Historically, physical models predicted the warp to be caused by a misaligned planet \citep{mouillet_97,augereau_01}, which was then observed \citep{lagrange_10}. Even though the dynamics in a debris disc is very different from protoplanetary discs, where torques are communicated by pressure and damped by viscosity, this case underlines that planets can imprint their signature in the 3D structure of a disc.

The warp in the disc around $\beta$-Pic was measured by directly imaging the disc. In farther protoplanetary discs, this kind of observations is obviously much more difficult, even though it has been done in a few cases \citep[e.g. for the protoplanetary disc 114-426 in Orion,][]{miotello_13}. However, \citet{rosenfeld_12} have recently shown that it is becoming possible to potentially measure warps with high precision by looking at the emission line profiles of the gaseous component. They have applied this method to the closest transition disc ever detected, TW Hydrae. We will thoroughly describe the system and its observations in Section \ref{sec:tw_hya}. \citet{rosenfeld_12} observed the disc with the Atacama Large Millimiter/submillimiter Array (ALMA) in the $^{12}$CO $J=2-1$ and $^{12}$CO $J=3-2$ emission lines at 1.3 and 0.87 mm, respectively. In both bands, they detect an enhanced emission in the inner regions of the disc. In this paper, they propose three theoretical models to explain such observation: a hotter inner disc, a non-Keplerian velocity field due to possible magnetic pressure effects \citep{shu_07}, and a warp, where the disc has Keplerian motions ($v_\phi=v_{\rm K}$), but the variation of the inclination $i$ with radius might be enhancing the projected radial velocity. Note that \citet{rosenfeld_13} just showed that the same effect could be due to fast radial inflows. In this paper we do not want to compare all the possible models. We consider the warp to be the most plausible, since other measurements in scattered light have enlightened an azimuthal asymmetry which is compatible with being caused by a warp \citep{roberge_05,debes_13}. We use this system as a template, in order to show the applicability of our models (see below), which relate the amount of warping in the disc with the dynamical properties of a possible misaligned central planet.

To summarise: in this paper we address two main issues. Firstly, we provide a simple prescription to relate the amount of warping of a circumbinary disc to the properties of the two central objects. The binary can be composed by two stars (stars or brown dwarfs) or by a star and a massive planet. In this way, we will have a simple observational feature to infer the presence of a planet in protoplanetary and transition discs. We will be very general on this first problem, by exploring a parameter space that is suitable for a secondary companion with a mass ranging from a stellar to a planetary one. The central object can be either a star or a brown dwarf \citep[since recent observations have started to focus on discs around brown dwarfs, see e.g.][]{ricci_12,ricci_13}. Secondly, we apply this simple model to the case of TW Hya, and see whether the observations by \citet{rosenfeld_12} are compatible with a massive planet in the disc's cavity. This will be an example of how this technique could be used, since measurements of spatially resolved line profiles in similar discs are going to become more and more accessible, due to ALMA reaching its full potential in the next few years. We also predict the amount of warping in LkCa 15 and T Cha, in case the observed companion were misaligned with respect to the disc.

In Section \ref{sec:theory} we briefly present an overview of the theory on warp propagation in protoplanetary discs, and we illustrate the models we use, that are mostly based on \citet{facchini13_1}. In Section \ref{sec:relations} we explore the parameter space, and we give some simple prescription where possible. Where it is not, we show plots where the warping of the disc is related to the properties of both the central binary and the disc. In Section \ref{sec:tw_hya} we address the case of TW Hya specifically, and we briefly focus on LkCa 15 and T Cha. Finally, in Section \ref{sec:disc} we discuss the results and draw our conclusions.

\section{Warp propagation in protoplanetary circumbinary discs}
\label{sec:theory}

In order to describe the warping of a protoplanetary disc via simple physical parameters, we need to find the steady-state configurations of the disc when it is affected by an external torque. In this case, we will focus on the secular torque caused by a central binary misaligned with respect to the outer regions of the disc. This description allows us to model both a proper circumbinary disc, where the binary is formed by two stars, and a transition disc where a misaligned massive planet is clearing out the dusty and gaseous component in the inner region of a protoplanetary disc. We refer to the work by \citet{facchini13_1} \citepalias[hereafter][]{facchini13_1} and \citet{facchini13_2}, where a more complete overview is reported.

Warp propagation in accretion discs occurs in two regimes. A first one, where the warp propagates via bending waves, and a second one, where it propagates diffusively. The threshold between the two regimes is obtained by comparing the dimensionless viscosity $\alpha$ \citep{shakura73} with the scale-height of the disc $H$. When $\alpha<H/R$, where $H/R$ is the aspect ratio, the warp evolves via wave equations \citep{papaloizou_lin95,lubow_ogilvie2000}, while when $\alpha>H/R$ the evolution is diffusive \citep{papaloizou_pringle83,pringle92}. 

For protoplanetary discs, typical values of $\alpha$ and $H/R$ are $\sim 10^{-4}-10^{-2}$ and $0.01-0.1$, respectively \citep[e.g.][]{hartmann_98,hueso_05}. Therefore, warps in protoplanetary discs are more likely to evolve via bending waves. In Section \ref{subsec:equations} we present the linearised equations for warp propagation in this case.

\subsection{Brief theory of warp propagation}
\label{subsec:equations}
In the thin disc approximation, the dynamics of almost Keplerian discs can be described in terms of some relevant physical quantities in the disc's local plane. We define the angular momentum per unit area ${\bf L}(R)=\Sigma R^2 \Omega {\bf l}$, where $\Sigma(R)$ is the surface density, $\Omega(R)$ the angular velocity and ${\bf l}(R)$ the specific angular momentum. If the disc is nearly Keplerian and non-self-gravitating, the linearised wave equations describing the warp evolution are \citep{lubow_ogilvie2000}:

\begin{equation}
\label{eq:wave_l_real}
\Sigma R^2\Omega\frac{\partial {\bf l}}{\partial t}=\frac{1}{R}\frac{\partial {\bf G}}{\partial R}+{\bf T},
\end{equation}
and
\begin{equation}
\label{eq:wave_g_real}
\frac{\partial{\bf G}}{\partial t}+\left(\frac{\kappa^2-\Omega^2}{\Omega^2}\right)\frac{\Omega}{2}{\bf e}_z\times{\bf G} +\alpha\Omega{\bf G}=\Sigma R^3\Omega\frac{c_{\mathrm{s}}^2}{4}\frac{\partial{\bf l}}{\partial R},
\end{equation}
where
\begin{equation}
\label{eq:external_torque}
{\bf T}=-\Sigma R^2\Omega\left(\frac{\Omega_z^2-\Omega^2}{\Omega^2}\right)\frac{\Omega}{2}{\bf e}_z\times{\bf l},
\end{equation}
The unit vector ${\bf e}_z$ is parallel to the $z$-axis, along which the gravitational potential is cylindrically symmetric. In this case it will be perpendicular to the binary plane. The other quantities are the horizontal internal torque $2\pi {\bf G}$, the external torque ${\bf T}$ (i.e. the forcing term), the epicyclic frequency $\kappa$, the vertical oscillation frequency $\Omega_{\rm z}$ and the sound speed $c_{\rm s}$. The sound speed is related to the scale-height of the disc via the relation $H=c_{\rm s}/\Omega$. All the frequencies $\Omega$, $\kappa$ and $\Omega_z$ can be computed from the form of the gravitational potential (see below equations (\ref{eq:omegaz2}) and (\ref{eq:kappa2})). These equations are derived by assuming a constant surface density $\Sigma$, since the sound crossing time is orders of magnitude lower than the viscous time in the bending wave regime \citepalias{facchini13_1}. They have the typical form of a wave equation for the angular momentum ${\bf L}$, more specifically for the angular momentum ${\bf l}$ since $\Sigma$ is considered constant. We can relate the specific angular momentum to more intuitive variables by defining ${\bf l}(R) = (\cos\gamma\sin\beta,\sin\gamma\sin\beta,\cos\beta)$. The angle $\beta$ defines the tilting with respect to the $z$-axis, whereas the angle $\gamma$ defines the orientation of the tilt with respect to an arbitrary axis, perpendicular to $z$. When $\partial_R\beta\neq 0$ we say that the disc is warped, when $\partial_R\gamma\neq 0$ we say that the disc is twisted. Heuristically, the equations reported above describe the fact that the disc can be warped and twisted when the epicyclic frequency and/or the vertical oscillation frequency differ from the angular frequency. This happens when the gravitational potential is not spherically symmetric, i.e. when the gravitational potential of the central star is perturbed. In the case of protoplanetary discs, the perturber could be a stellar flyby \citep[e.g.][]{clarke_93} or a stellar companion \citep[e.g.][]{lubow_ogilvie01,foucart_lai_13}.

By solving equations (\ref{eq:wave_l_real}) and (\ref{eq:wave_g_real}) we can describe the wave-like evolution of the warp, and look for the steady-state configuration of the disc. In order to do so, we need to evaluate the three frequencies included in the equations, and we need to choose a disc model, i.e. an analytical profile for the surface density $\Sigma$ and the sound speed $c_{\rm s}$.

In this paper we focus on a gravitational potential generated by a central binary. The mass ratio of the two objects can reach extreme values, therefore in principle this same model can be used to describe a binary formed by a central star and an orbiting massive planet. We follow the same procedure as in \citetalias{facchini13_1}. We consider two stars $M_1$ and $M_2$ co-orbiting on circular orbits. The binary system can be described via the standard two body variables $M = M_1+M_2$, $\eta = M_1M_2/M^2$ and $a$, where $a$ is the distance between the two central objects. The gravitational potential is computed by considering the prominent secular term only, i.e. the time independent term \citep[][]{nixon2011,foucart_lai_13}. \citet{bate_00} have shown that time dependent terms introduce negligible effects, as confirmed numerically by \citetalias{facchini13_1}. We then expand the secular term in powers of $a/R$ and $z/R$ up to second order. This expansion requires  $a/R\ll1$ and  $z/R\ll1$, conditions that are satisfied for a thin circumbinary disc. We obtain the following gravitational potential $\Phi$:

\begin{equation}
\label{eq:phi_fin}
\nonumber\Phi(R,z)=-\frac{GM}{R}-\frac{GM\eta a^2}{4R^3} + \frac{GMz^2}{2R^3}+\frac{9}{8}\frac{GM\eta a^2z^2}{R^5}.
\end{equation}
By using this relation, we derive the following ratios, to be used to solve equations (\ref{eq:wave_l_real})-(\ref{eq:external_torque}):

\begin{equation}
\label{eq:omegaz2}
\frac{\Omega_z^2-\Omega^2}{\Omega^2}=\frac{3}{2}\frac{\eta a^2}{R^2},
\end{equation}
\begin{equation}
\label{eq:kappa2}
\frac{\kappa^2-\Omega^2}{\Omega^2}=-\frac{3}{2}\frac{\eta a^2}{R^2}.
\end{equation}

Finally, we use a prescription for the disc where both the surface density and the sound speed depend on the radial coordinate via a simple power law; respectively, $\Sigma \propto R^{-p}$ and $c_{\rm s} \propto R^{-q}$. The radial extent of the disc is limited by an inner and an outer edge, $R_{\rm in}$ and $R_{\rm out}$. From now on the subscript \emph{in} refers to quantities evaluated at the inner edge of the disc.

By using these analytical prescriptions for the gravitational potential and for the disc, we can follow the warp evolution in circumbinary discs.

\subsection{Steady-state solutions}

It is well known that warped discs that evolve via bending waves reach a steady-state configuration on a timescale of the order of a few sound crossing times, since the waves propagate with a velocity of $c_{\rm s}/2$ \citep{nelson_pap99,LOP02}. Two ways have been recently used to find the steady-state solutions for both the warp and the twist. \citet{foucart_lai_13} integrated semi-analytically equations (\ref{eq:wave_l_real})-(\ref{eq:external_torque}) by setting to 0 the time dependent terms ($\partial/\partial t=0$), whereas \citetalias{facchini13_1} followed the time dependent evolution, until the disc relaxed to its steady-state configuration. Here we use the same method as described by \citetalias{facchini13_1}. This method has been tested via full 3D SPH (Smoothed Particle Hydrodynamics) simulations using the \textsc{phantom} code, which has shown good agreement with the 1D analytic theory in the diffusive regime \citep{lodato_price10}. \citetalias{facchini13_1} found a good agreement in the bending-wave regime, both in the diffusive limit ($\alpha > H/R$) and in the inviscid limit ($\alpha \lesssim 0.01$). With this method, we are able to obtain the steady-state solution of a disc misaligned with respect to the central binary's plane.

The warped structure will be expressed in terms of the angle $\beta(R)$. The parameter $\beta_\infty$ will indicate the angle between the outer disc's plane and the binary. We recall that we are focusing on the linear regime only, since it is the only regime for which an analytic theory has been developed. \citetalias{facchini13_1} have shown that for $\beta_\infty \gtrsim 40^\circ$ the evolution becomes non-linear, and the disc can break. In this work we do not consider this regime.

We stress that the steady-state solution will depend on four parameters. The two exponents $p$ and $q$ from the disc model, and two dimensionless parameters: $\hat{\alpha}=\alpha/(H_{\rm in}/R_{\rm in})$ and $\chi$ \citepalias[cfr. equations (33)-(36) in][]{facchini13_1}, where the parameter $\chi$ is defined as:

\begin{equation}
\chi=\frac{3}{4}\eta\frac{(a/R_{\rm in})^2}{H_{\rm in}/R_{\rm in}}.
\label{eq:defchi}
\end{equation}
Physically, these two ratios indicate that the dynamics of the disc is regulated by two physical ingredients: the relative importance of viscous over pressure forces, and the relative importance of the external gravitational torque over internal pressure stresses.

\section{Results}
\label{sec:relations}

We have performed 1D simulations using the model described in Section \ref{sec:theory}. Our aim is to explore the 2D $\hat{\alpha}$-$\chi$ parameter space for some typical values of $p$ and $q$, in order to find a simple prescription for the dependence of the amount of warping in a disc with respect to these two parameters. \citet{foucart_lai_13} have already found this kind of prescription, but focusing on the case where $q$ is a function of $p$ only. In this section we will compare our results with theirs. Moreover, our method allows us to explore a region of parameter space that was not considered by \citet{foucart_lai_13}. They focus on cases where $R_{\rm in} > R_{\rm warp}$. The quantity $R_{\rm warp}$ can be defined by the following relation:

\begin{equation}
R_{\rm warp} \approx a \left( \frac{3\alpha\eta} {2(H_{\rm in}/R_{\rm in})^2} \right)^{1/2} = R_{\rm in} (2\hat{\alpha}\chi)^{1/2},
\end{equation}
and it indicates the scale radius where the warp becomes prominent. In terms of limits, it means that for $R\ll R_{\rm warp}$ the disc is aligned with the binary plane, whereas for $R \gg R_{\rm warp}$ the disc is inclined by a same angle $\beta_{\infty}$. In this paper, we explore regions of parameter space where $R_{\rm in}\approx R_{\rm warp}$, e.g. by exploring a highly viscous regime or a regime where the binary torques are stronger. Note that $R_{\rm in}\approx R_{\rm warp}$ when $\hat{\alpha}\chi\approx 1$.

\begin{table}
\centering
\begin{tabular}{cccc}
\hline
$p$		& $q$	& $K$	& $C$	 \\
\hline
0.50		& 0.50	& 4.1045	& 0.00991 \\
1.00		& 0.50	& 23.891	& 0.02380 \\
0.93		& 0.26	& 2.5072	& 0.12766 \\
0.72		& 0.15	& 0.7256	& 0.37397 \\
0.72		& 0.30	& 1.8249	& 0.18128 \\
1.00		& 0.40	& 9.7410	& 0.04472 \\
\hline
\end{tabular}
\caption{Obtained values for $K$ and $C$ for different choices of density and sound speed power-law indexes $p$ and $q$.  The parameters were deduced by fitting the fractional amplitude of the warping $\Delta\beta/\beta_\infty$ obtained via 1D simulations. These simulations explore a wide parameter space in $\chi$ and $\hat{\alpha}$. The fitting function is given by equation (\ref{eq:deltabeta}). The last four choices of $p$ and $q$ represent physical systems (TW Hya, LkCa 15 and T Cha, respectively) that are discussed in Section \ref{sec:obs}. Comparisons between outputs of the simulations and fitting curves are displayed in Figs. \ref{fig:tw_m080}-\ref{fig:lkca15}.}
\label{tab:values}
\end{table}

The simulations are performed for different combinations of the parameters $p$ and $q$. Firstly we have considered typical density and sound speed profiles. We have used: $(p,q)=(1.00,0.50)$ and $(0.50,0.50)$. We have then chosen parameters that were deduced in best fits of observed systems. We focus on the three systems we discuss in Section \ref{sec:obs}: TW Hya, LkCa15 and T Cha. The respective values for $(p,q)$ are: $(0.93,0.26)$, $(0.72,0.15-0.30)$ and $(1.00,0.40)$. The references for these values are summarised in Table \ref{tab:discs}. For every couple of $p$ and $q$ values, we run simulations for $\hat{\alpha}=[0.00,3.00]$, sampled at every $\Delta \hat{\alpha} = 0.1$, and $a/R_{\rm in} =[0.05,1.00]$, sampled at every $\Delta (a/R_{\rm in})=0.05$ for $\eta=0.25$. This is equivalent to sampling $\chi$ between $4.69\cdot10^{-3}$ and $1.875$. Note again that the two ranges allow us to explore extreme conditions, such as two equally massive stars orbiting at the very inner edge of a disc. Obviously this kind of conditions will never be observed in relaxed environments, but it could be observed in short transients in dynamically interactive regions.

\begin{figure*}
\begin{center}
\includegraphics[width=.95\columnwidth]{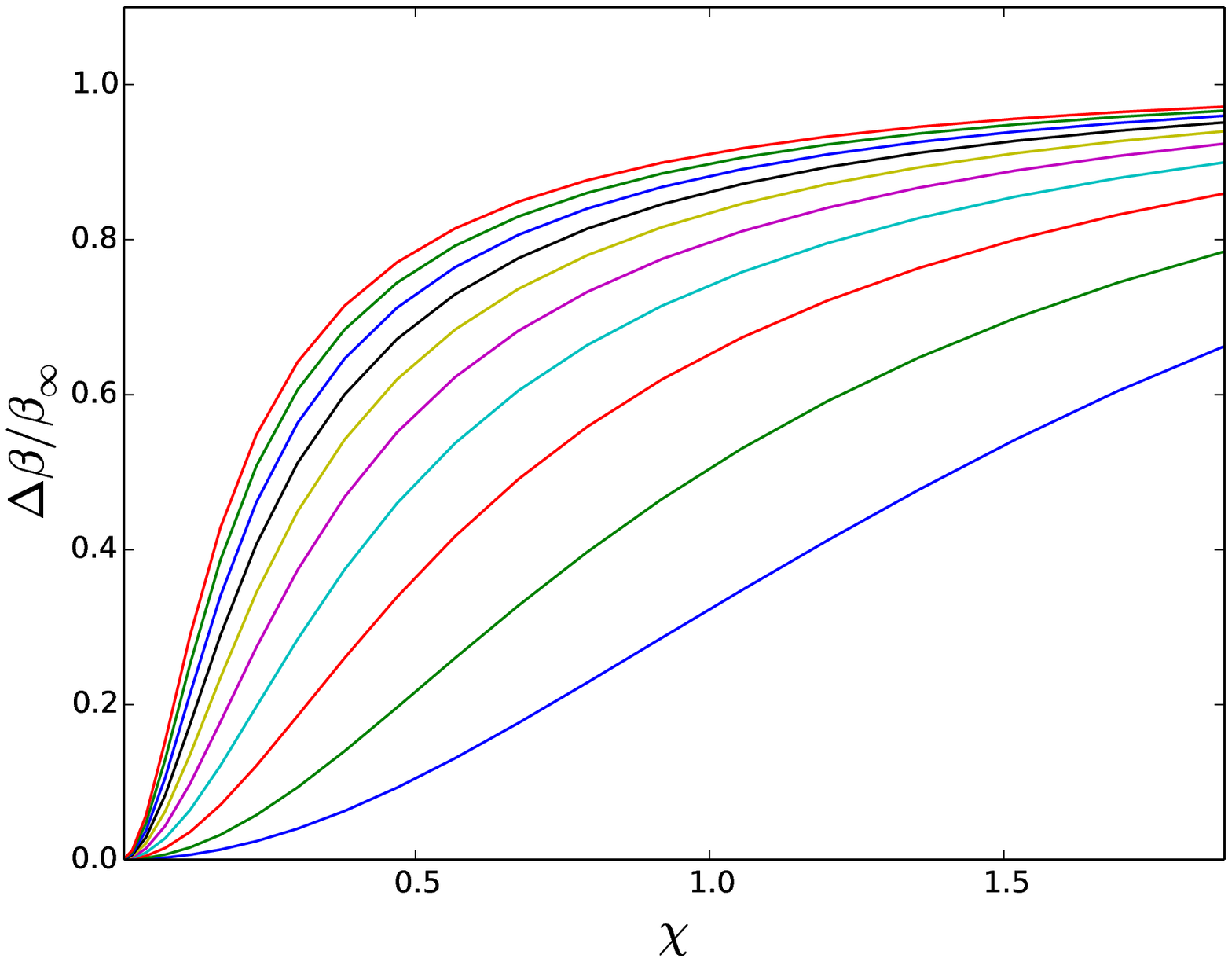}
\includegraphics[width=.95\columnwidth]{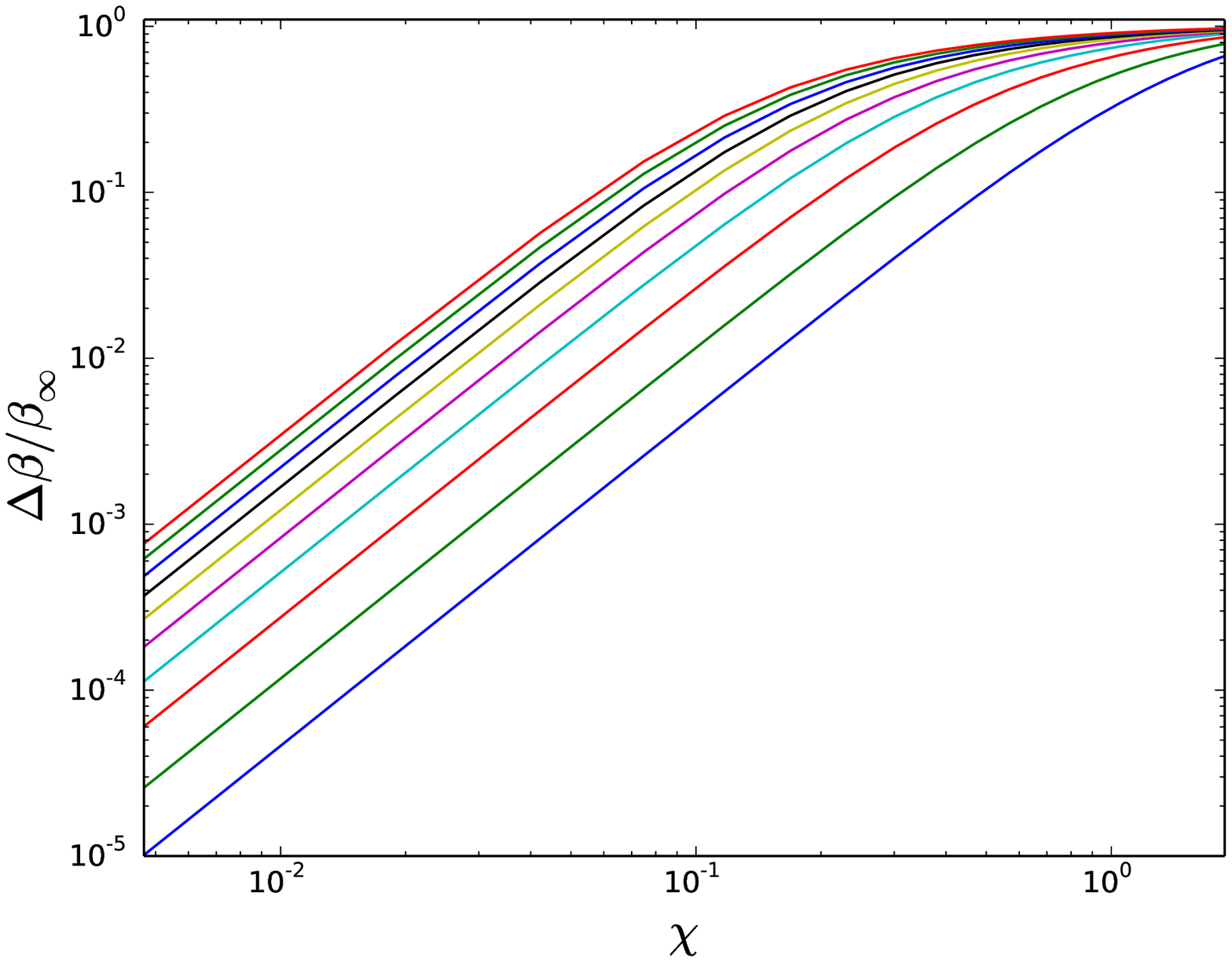}\\
\includegraphics[width=.95\columnwidth]{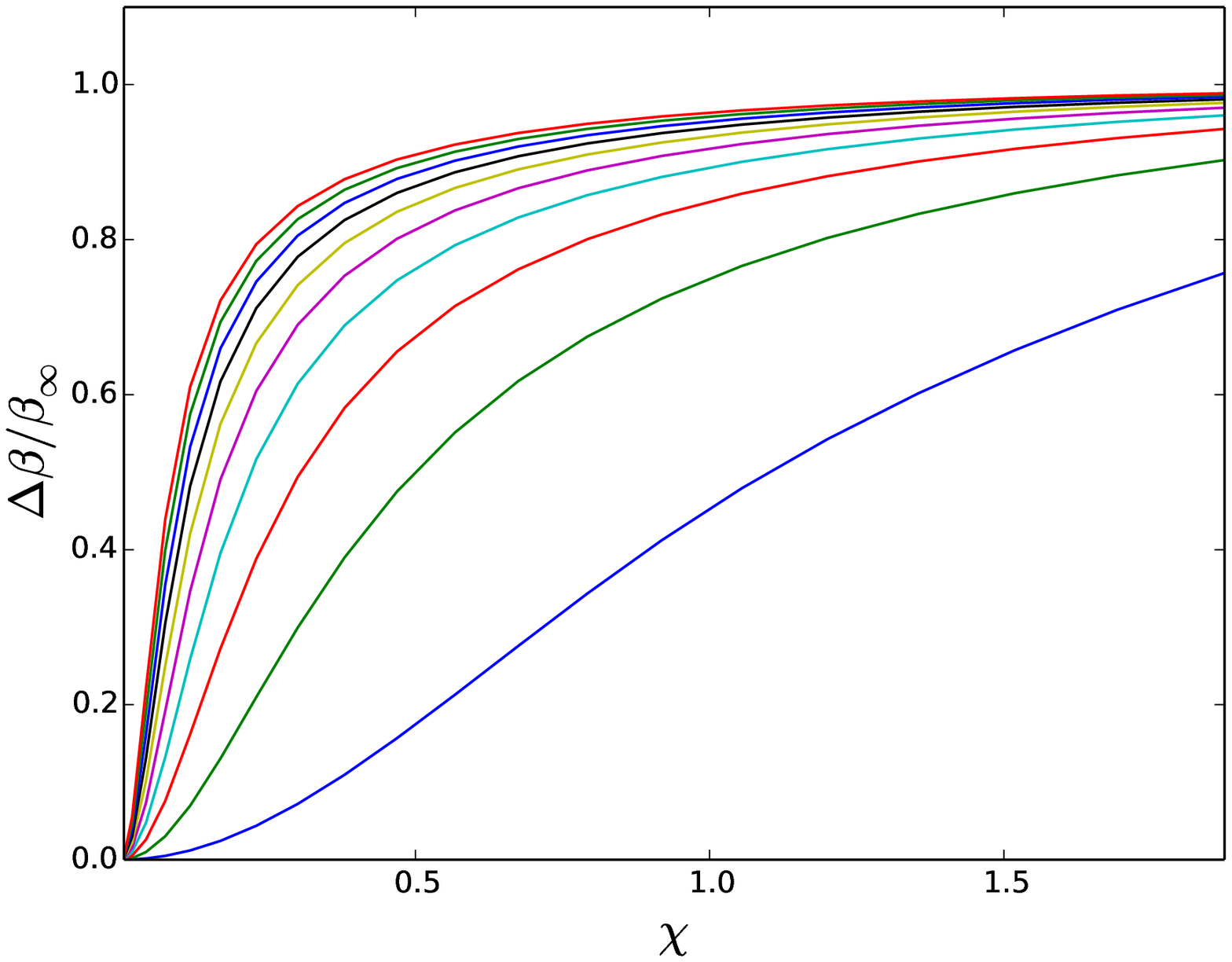}
\includegraphics[width=.95\columnwidth]{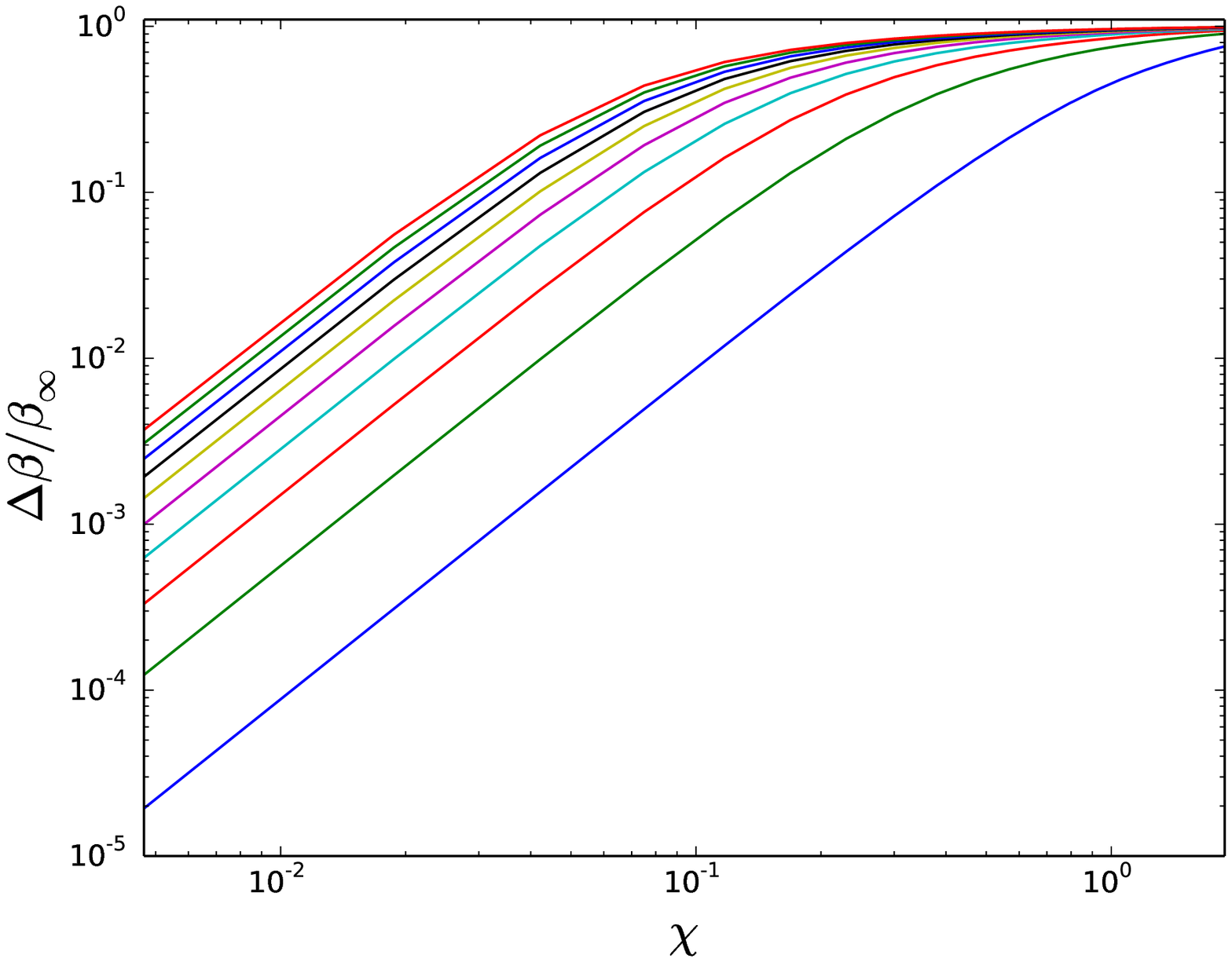}\\
\includegraphics[width=.95\columnwidth]{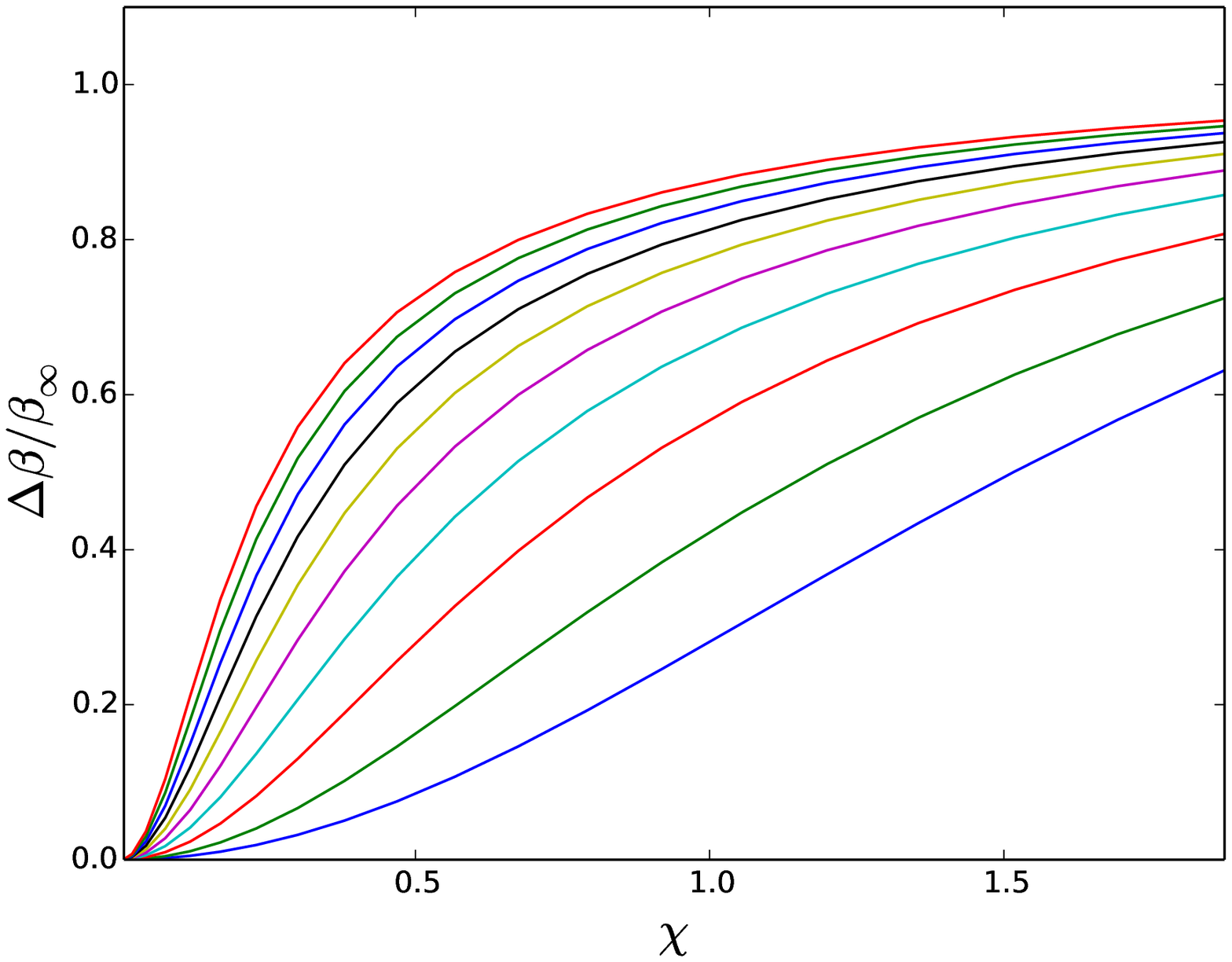}
\includegraphics[width=.95\columnwidth]{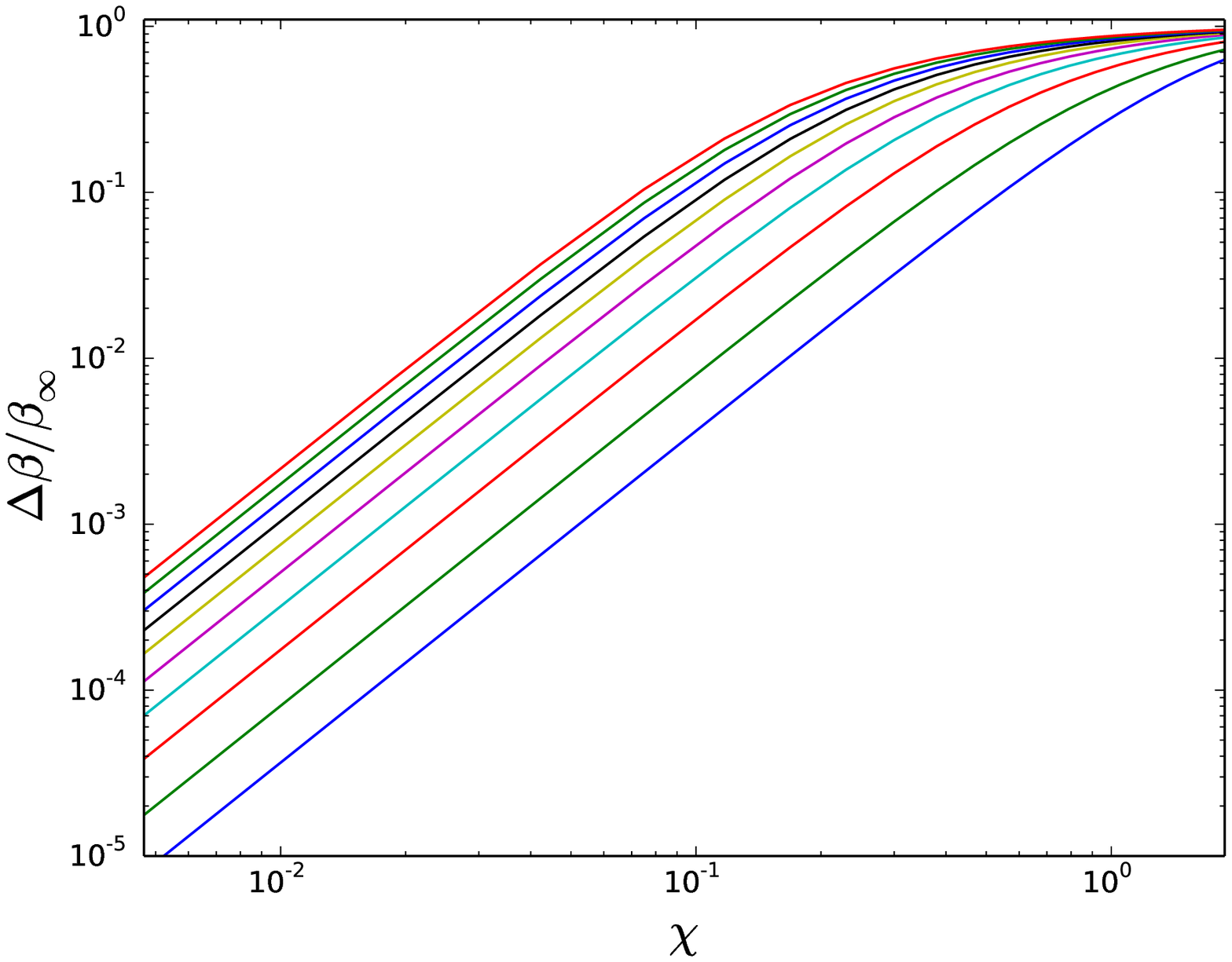}
\end{center}
\caption{Warping of the steady-state solution of a circumbinary disc as a function of $\chi$ for three different values of the couple $(p,q)=(0.50,0.50)$, $(1.00,0.50)$ and $(0.93,0.26)$, from top to bottom respectively. The right panels are the log-log scaled versions of the left panels. Each line represents a simulation at a different value of viscosity: from $\hat{\alpha}=0.1$ (blue line at the bottom of each plot) to $\hat{\alpha}=2.8$, sampled every $\Delta\hat{\alpha}=0.03$. The amount of warping depends on $\chi$ as a power law when the warping is small. In particular, $\Delta\beta / \beta_{\infty} \propto \chi^2$. These results confirm relations (16), (20) and (21) by \citet{foucart_lai_13}. As the warp grows the dependence on $\chi$ deviates from a simple power law, as it flattens so that $\Delta\beta / \beta_\infty$ saturates to $1$ when $\chi$ is large. The simple quadratic dependence on $\chi$ is only valid as long as $\Delta\beta / \beta_\infty \lesssim 0.1$. }
\label{fig:tilt_chi}
\end{figure*}

\begin{figure*}
\begin{center}
\includegraphics[width=.95\columnwidth]{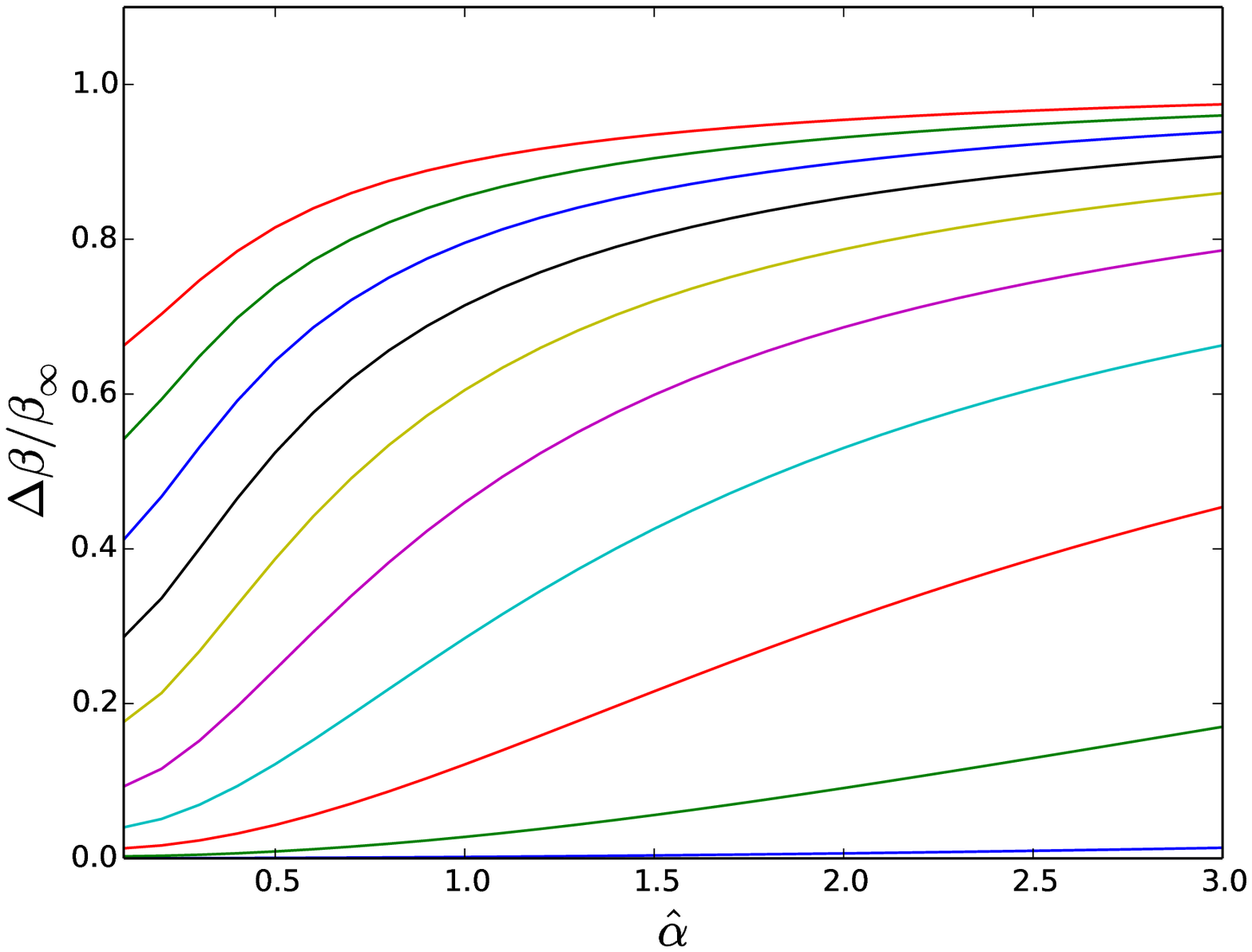}
\includegraphics[width=.95\columnwidth]{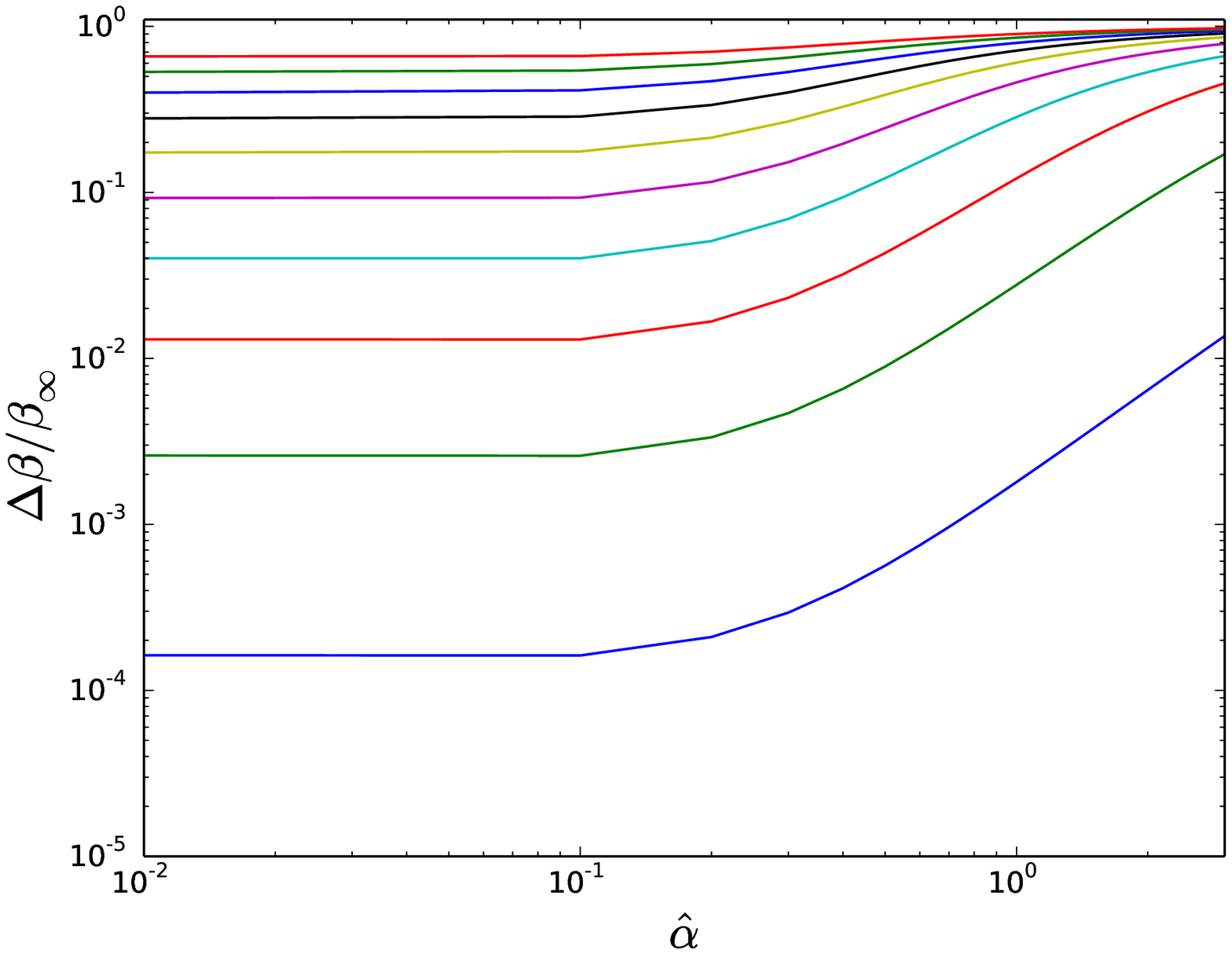}\\
\includegraphics[width=.95\columnwidth]{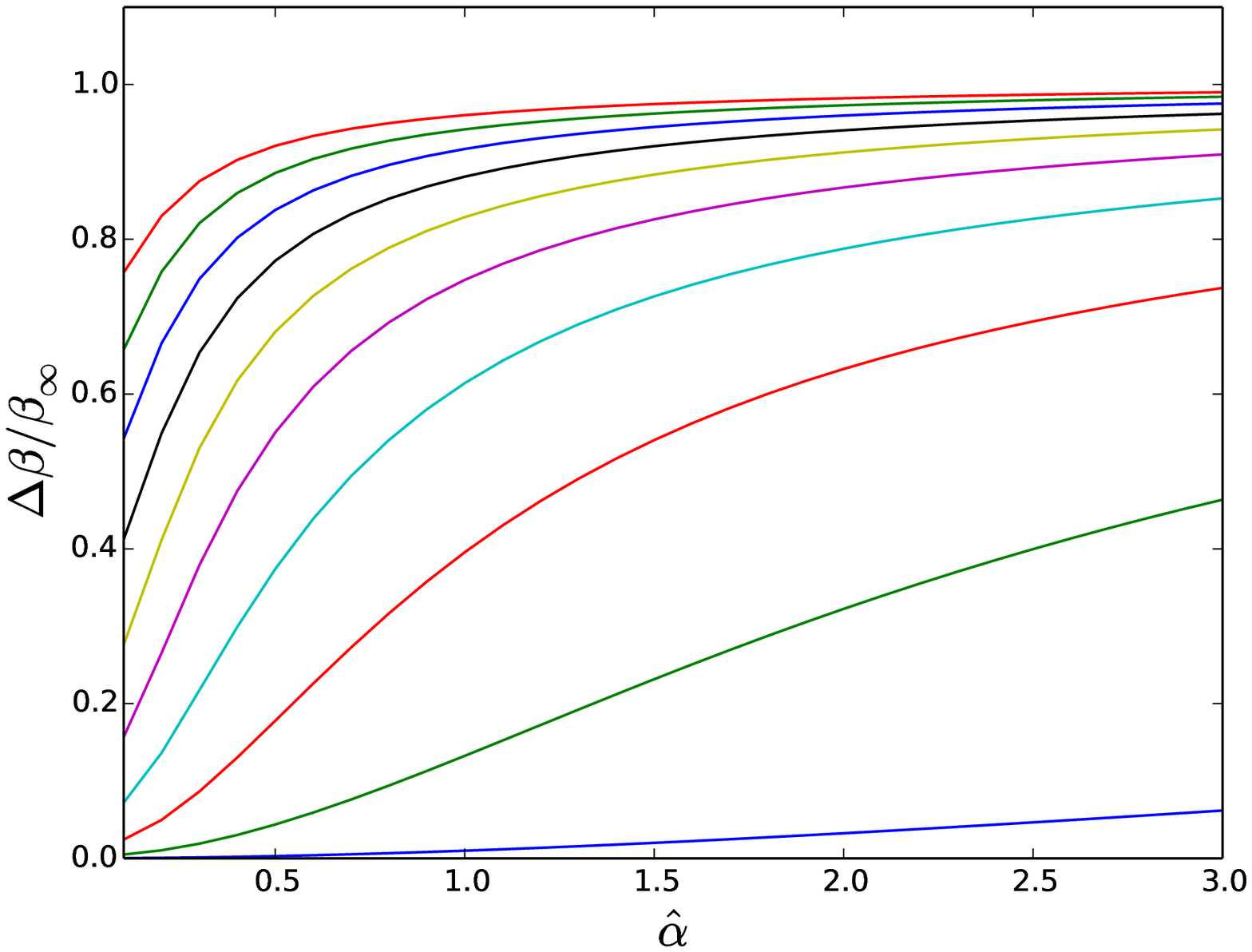}
\includegraphics[width=.95\columnwidth]{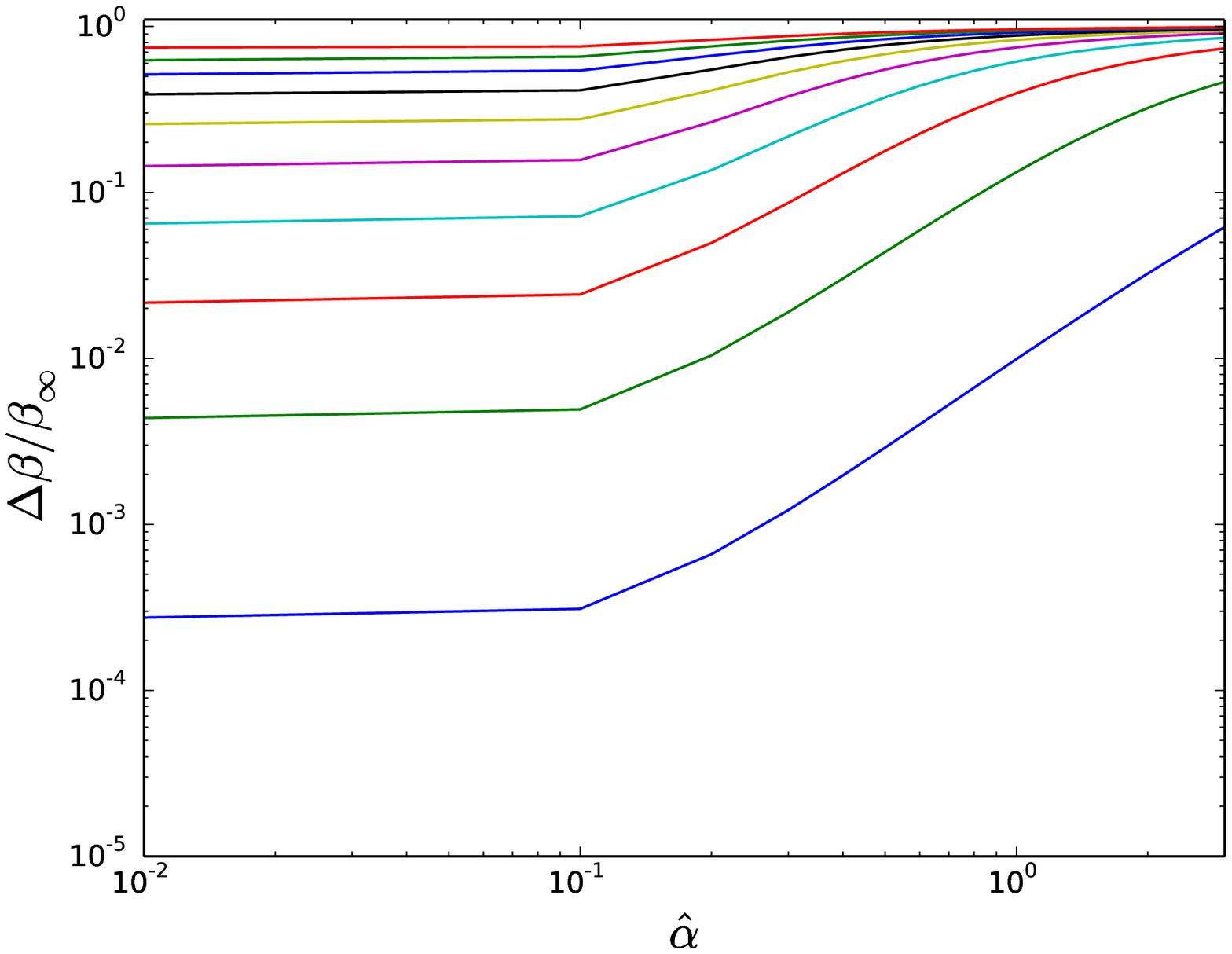}\\
\includegraphics[width=.95\columnwidth]{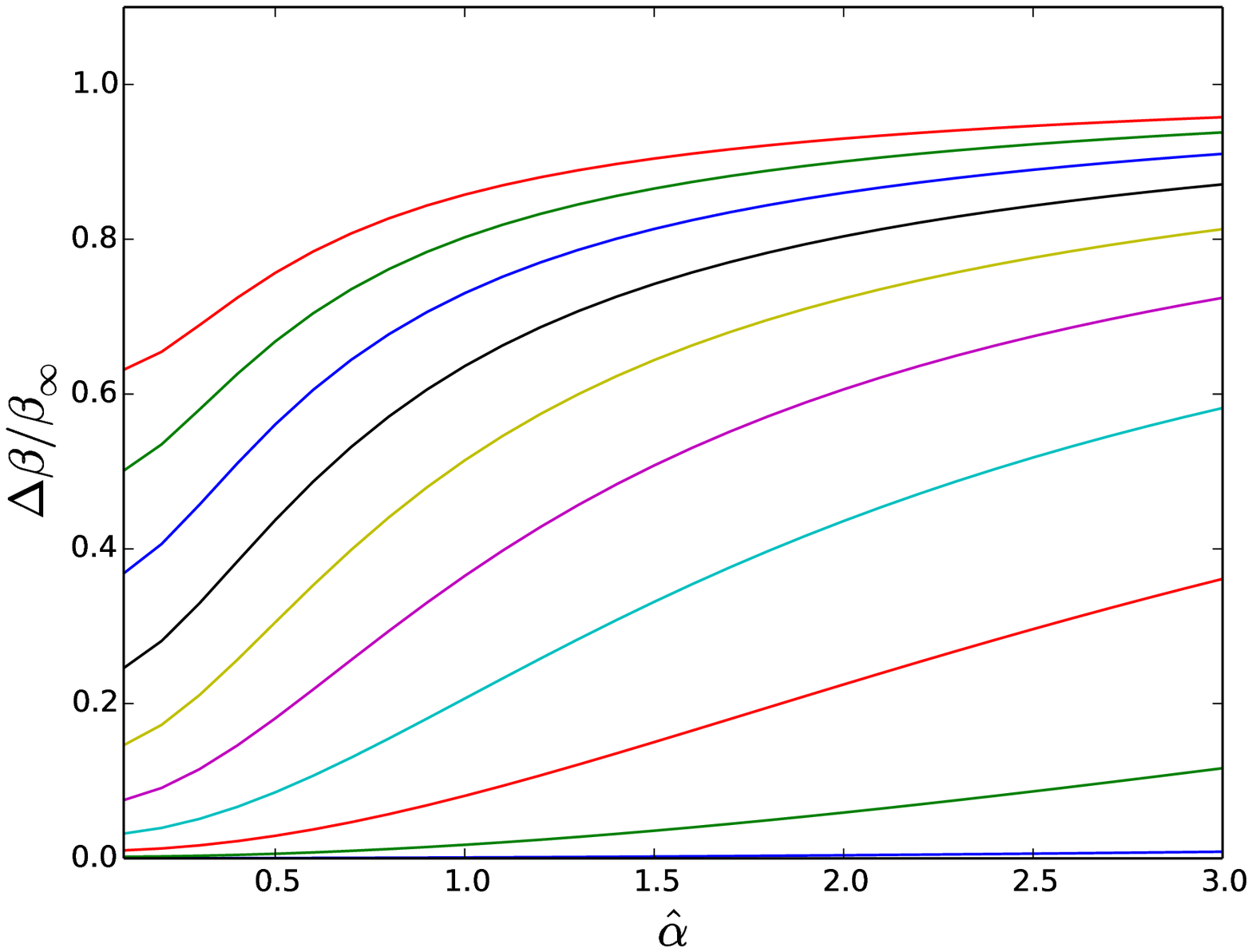}
\includegraphics[width=.95\columnwidth]{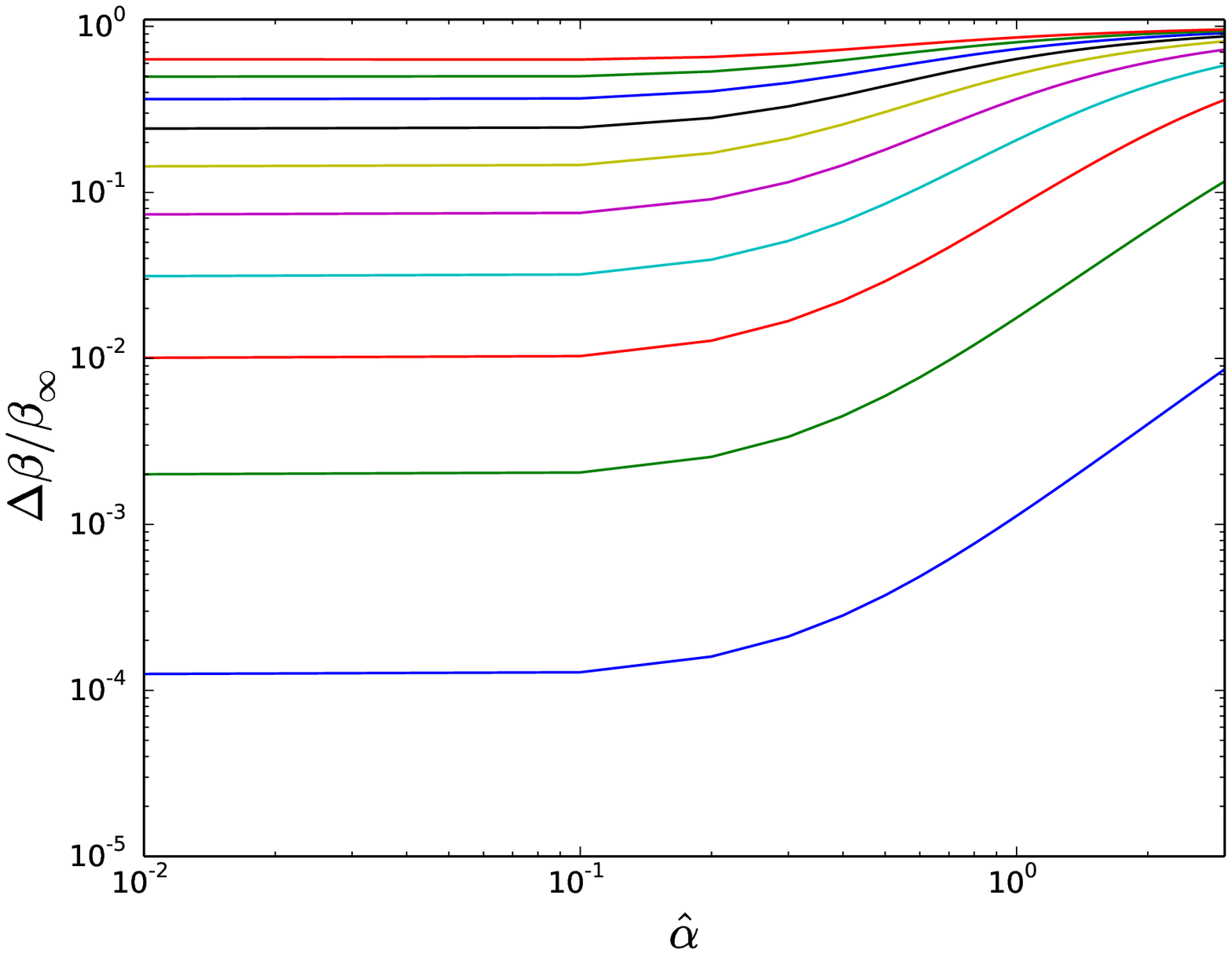}
\end{center}
\caption{Warping of the steady-state solution of a circumbinary disc as a function of $\hat{\alpha}=\alpha/(H_{\rm in}/H_{\rm in})$ for three different values of the couple $(p,q)=(0.50,0.50)$, $(1.00,0.50)$ and $(0.93,0.26)$, from top to bottom respectively. Each line represents a simulation at a different value of $a/R_{\rm in}$: from $a/R_{\rm in}=0.1$ (blue line at the bottom of each plot) to $a/R_{\rm in}=1.0$, sampled every $\Delta(a/R_{\rm in})=0.1$, for $\eta=0.25$. This corresponds to sampling $\chi$ from $\chi_1=0.01875$ to $\chi_{10}=1.875$, where $\chi_i=1.875*(a_i/R_{\rm in})^2$ and $a_i/R_{\rm in}=i*\Delta(a/R_{\rm in})$. We can observe three different regimes. At low $\hat{\alpha}$ the amount of warping reaches a horizontal asymptote. It then shows a power law trend in an intermediate regime ($\Delta\beta/\beta_\infty\propto\hat{\alpha}^2$). Roughly above the threshold $\Delta\beta / \beta_{\infty} = 0.1$ we obtain a non analytic relation again, as found for the parameter $\chi$.}
\label{fig:tilt_alpha}
\end{figure*}

The simulations are performed with $N=2001$ logarithmically distributed grid points. The simulations end when the perturbation coming from the inner edge of the disc reaches the outer edge $R_{\rm out}=500R_{\rm in}$. At this point, in the inner regions ($R<250R_{\rm in}$) the discs have already relaxed to their steady-state configuration even in the more viscous cases. The amount of warping in the disc is quantified by the quantity $\Delta\beta / \beta_{\infty} = (\beta_{\infty} - \beta_{\rm in})/\beta_{\infty}$. In these estimates, $\beta_{\infty} = \beta(R=120R_{\rm in})$. We consider an unwarped, misaligned initial condition, where $\beta(R) = \beta_\infty$ for every $R$. \citetalias{facchini13_1} have shown that the solution does not depend on the initial condition. Note that $\beta_\infty$ in computed far from $R_{\rm out}$ since we do not want to have any spurious effect due to the outer edge boundary condition. Since the warp develops in the inner regions only, ${\bf l}$ computed at $R=120R_{\rm in}$ is a very good approximation for ${\bf l}_{\rm out}$. We have used a zero-torque boundary condition both at the inner and at the outer edge of the disc ($\partial_R {\bf l}_{\rm in} = \partial_R {\bf l}_{\rm out} = 0$).

The results our reported in Figs. \ref{fig:tilt_chi}-\ref{fig:tilt_alpha}. All the figures illustrate the dependence of $\Delta\beta / \beta_{\infty}$, respectively as a function of $\chi$ and $\hat{\alpha}$. The plots are general, in the sense that they do not depend on the specific choice of $H_{\rm in}/R_{\rm in}$ and of $M_1$ and $M_2$, made in the simulations.

Let us start by looking at Fig. \ref{fig:tilt_chi}. It shows that the amount of warping depends on $\chi$ as a power law when the warping is small. In particular, $\Delta\beta / \beta_{\infty} \propto \chi^2$. These results confirm relations (16), (20) and (21) by \citet{foucart_lai_13} when $\sin\beta\sim\beta$. However, as the warp grows the dependence on $\chi$ deviates from a simple power law, as it flattens so that $\Delta\beta / \beta_{\infty}$ saturates to $1$ when $\chi$ is large, a result not predicted by the analytical model by  \citet{foucart_lai_13}. The simple quadratic dependence on $\chi$ is only valid as long as $\Delta\beta / \beta_{\infty} < 0.1$.

From Fig. \ref{fig:tilt_alpha} we obtain very similar results. From the plots we can observe three different regimes. At low $\hat{\alpha}$ the amount of warping reaches a horizontal asymptote. It then shows a power law trend in a intermediate regime. By fitting the curves in this regime we obtain $\Delta\beta / \beta_{\infty} \propto \hat{\alpha}^2$, as predicted by \citet{foucart_lai_13} by their equations (16) and (20). Roughly above the threshold $\Delta\beta / \beta_{\infty} = 0.1$ we obtain a non immediate relation again, as found for the parameter $\chi$. The horizontal asymptote was predicted by \citet{foucart_lai_13} again. They show that this asymptote is due to the non-Keplerian term in equation (\ref{eq:wave_g_real}). 

We can summarise these results by writing a simple prescription for $\Delta \beta / \beta_{\infty}$, whenever $R_{\rm warp} < R_{\rm in}$:

\begin{equation}
\label{eq:deltabeta}
\frac{\Delta \beta}{\beta_{\infty}} \approx K \chi^2 \left(\hat{\alpha}^2 + C \right).
\end{equation}

We fitted the results of the 1D simulations with this analytic prescription on the parameter space that has an associated degree of warping $\Delta\beta/\beta_\infty<0.1$. The obtained values for $K$ and $C$ in the explored set of $(p,q)$ is reported in Table \ref{tab:values}. When $R_{\rm warp} \sim R_{\rm in}$, this simple prescription breaks. Typical values for this regime can be easily obtained from Figs. \ref{fig:tilt_chi}-\ref{fig:tilt_alpha}. In Figs. \ref{fig:tw_m080}-\ref{fig:lkca15} we show the comparison of the results obtained from numerical simulations and the analytic prescription in some specific cases, that are described in Section \ref{sec:obs}. The agreement between the curves is remarkable when $\Delta\beta/\beta_\infty<0.1$. This result confirms that the power-law dependencies are valid for small warps only.

This simple prescription tells us that by measuring $\Delta\beta/\beta_{\infty}$ we can obtain significant information on $\chi$ and $\hat{\alpha}$. If other parameters are known, such as the orbital parameters of the two stars (for stellar binaries), the degeneracy between $\chi$ and $\hat{\alpha}$ can be broken. Therefore, by measuring the warping of a circumbinary disc, it is possible to estimate the disc viscosity, parametrised by $\alpha$. This topic is better discussed in Section \ref{sec:disc}.

\section{Application to observed systems}

\label{sec:obs}

In this section we apply the model to some observed circumbinary and transition discs. We do it for two reasons. Firstly, we want to show how the model described in Section \ref{sec:relations} can be applied to a physical system. Secondly, we want to verify whether the warp that was observed in one of these systems can be explained by invoking a perturbing planet that is simultaneously clearing out the inner region of the disc. We consider three cases: TW Hya, for which a warp has been inferred based on ALMA observations of the velocity field \citep{rosenfeld_12}, LkCa 15, and T Cha. These last two objects are transition disc for which a potential protoplanet has been imaged within the central cavity. 

\subsection{TW Hya}
\label{sec:tw_hya}

TW Hya is a protoplanetary disc $54\pm6$ pc away from Earth \citep[cfr. Hipparcos catalogue,][]{van_leeuwen_07}, first detected by \citet{rucinski_83} at infrared wavelengths. Due to its high emission at many wavelengths, it is one of the most studied protoplanetary discs. The disc is nearly face-on \citep[$i\approx 6^\circ-7^\circ$, where $i=0^\circ$ is face-on;][]{qi_04}. Note that this particular feature helped \citet{rosenfeld_12} to spatially resolve the projected velocity of the gas in the disc, and to look for any departure from the expected Keplerian velocity.

By modelling the SED (Spectral Energy Distribution) of TW Hya, \citet{calvet_02} claimed that the disc has a central cavity of 4 AU in radius, confirmed by \citet{andrews_12} again by looking at the SED. This result was then confirmed by interferometric measurements, both in the near-infrared band \citep{eisner_06} and in the millimetric band \citep{hughes_07}. Note however that \citet{ratzka_07} derived a cavity of $0.7$ AU in radius by modelling their mid-infrared interferometric data. A few papers have tried to reconcile these two results by inferring a central cavity of 4 AU, with an inner source of emission coming from within the cavity \citep{akeson_11,arnold_12}. Since the paper by \citet{calvet_02}, many observers proposed a giant planet to be the cause of such an optically thin regime in the inner regions. Recently, \citet{evans_12} set an observational upper limit of $14M_{\rm J}$ for the purported planet/companion from a near-infrared aperture masking interferometry survey. However, other hypotheses have been proposed: interested readers are referred to \citet{gorti_11}, \citet{pascucci_11} and \citet{alexander_13} for a discussion on the physical origin of the central hole, where internal photoevaporation and a giant planet hypotheses are considered and compared. The presence of a central planet is anyway highly plausible. It is very interesting that a natural explanation of the warping of the disc is therefore a misaligned planet, which induces a warp in the inner region of the disc, and simultaneously clears out the inner cavity. In this section we verify whether this picture is compatible with observations.

\begin{figure*}
\begin{center}
\includegraphics[width=.95\columnwidth]{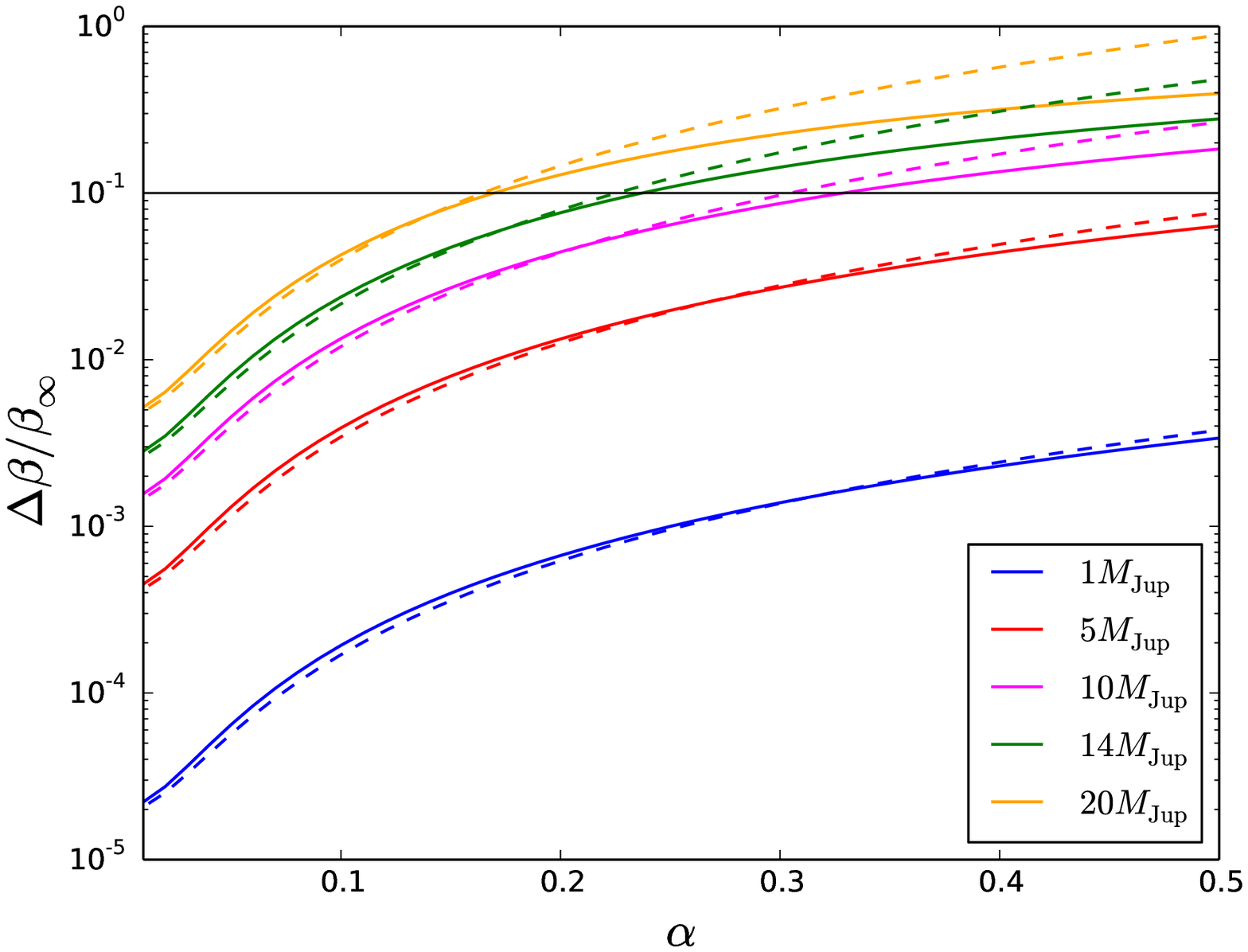}
\includegraphics[width=.95\columnwidth]{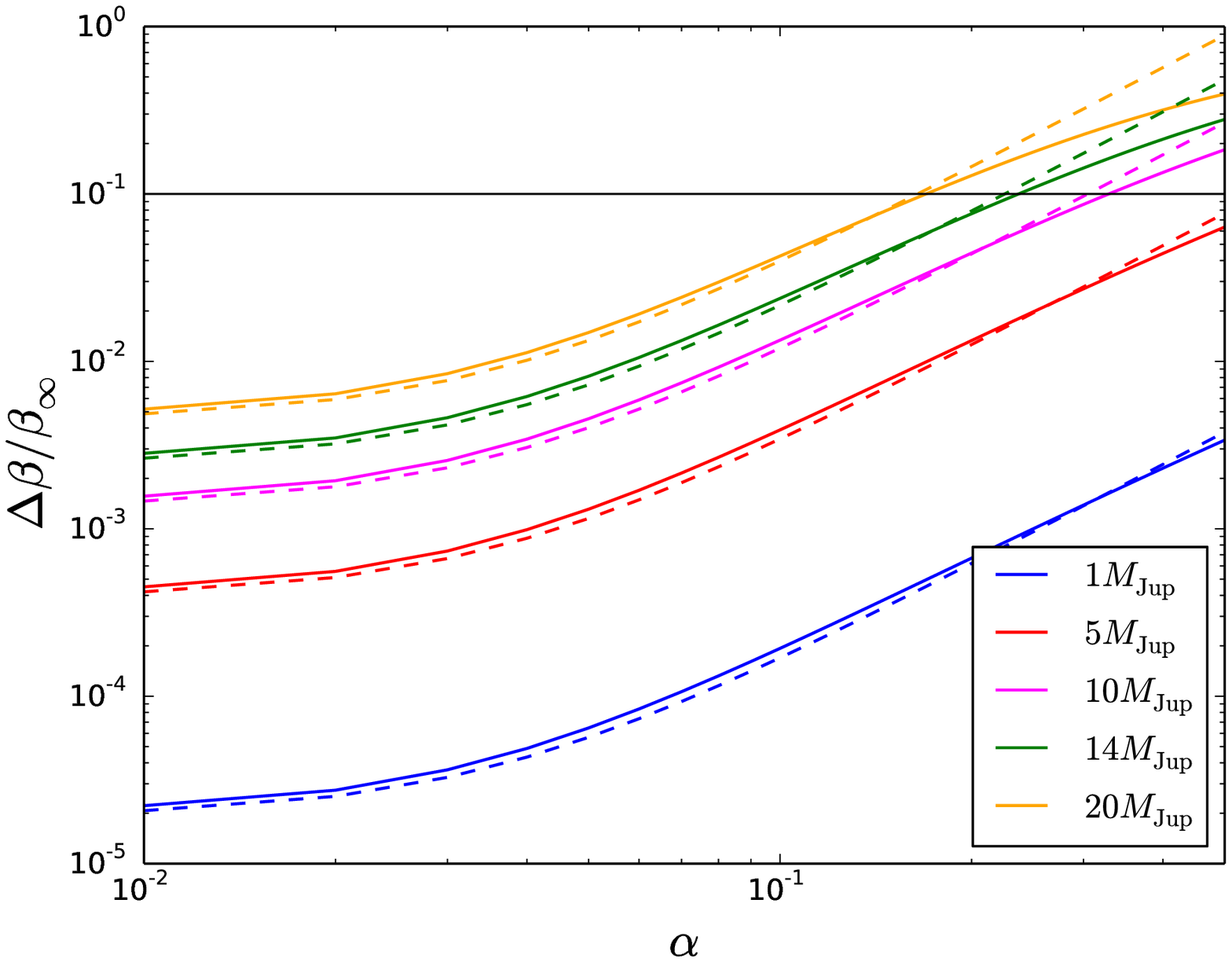} \\
\includegraphics[width=.95\columnwidth]{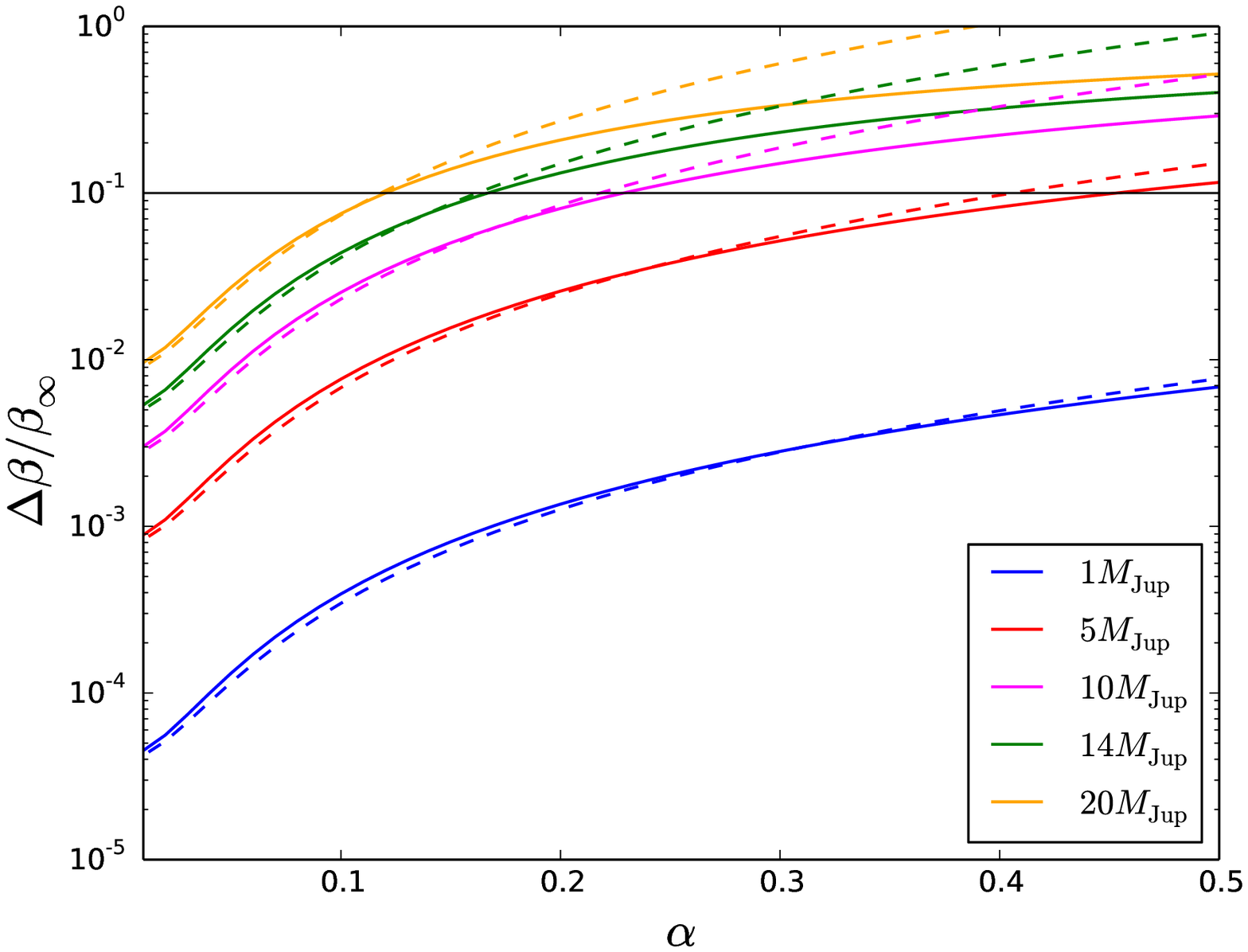}
\includegraphics[width=.95\columnwidth]{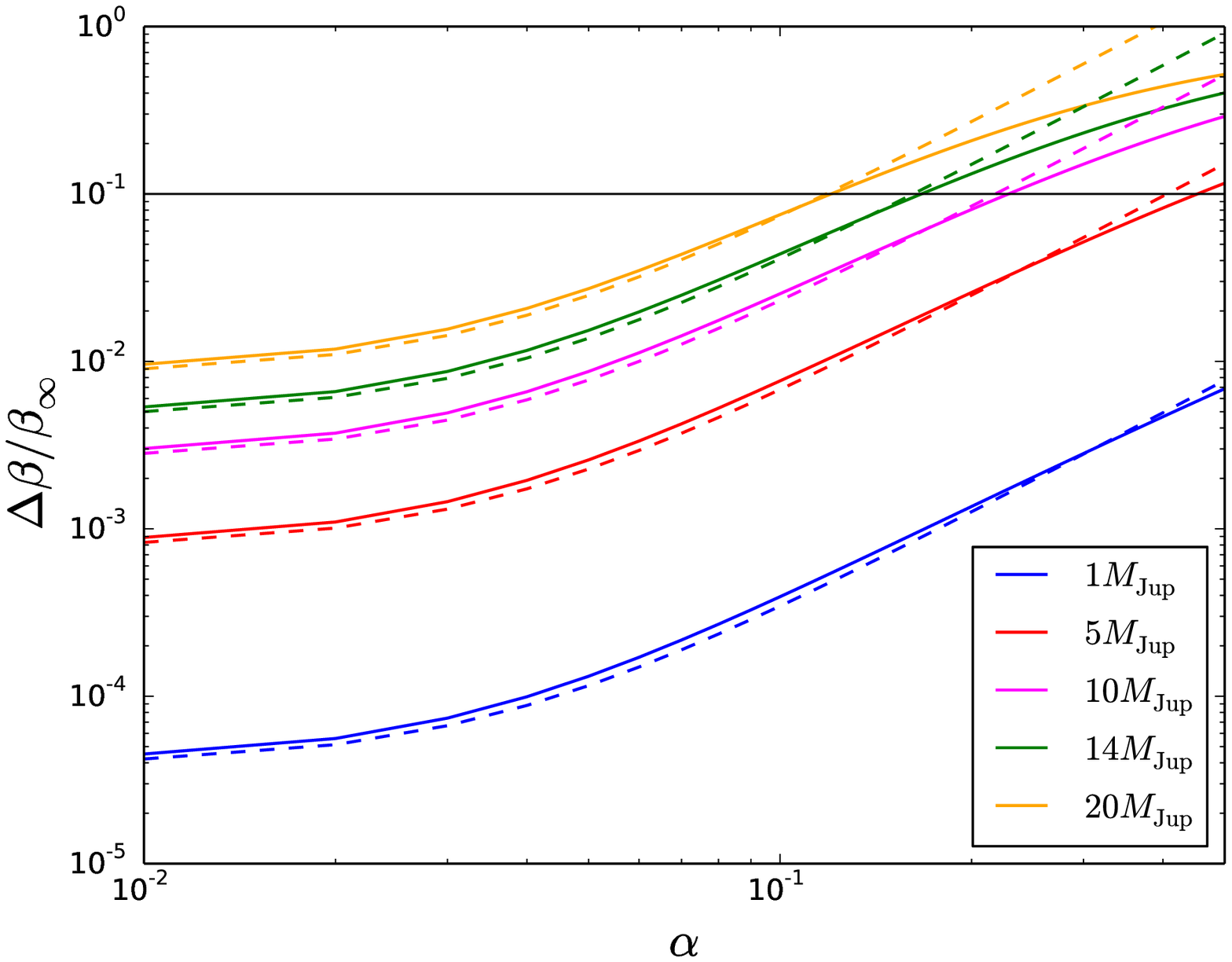}
\end{center}
\caption{Top two panels: $\Delta\beta/\beta_\infty$ as a function of viscosity for different planets masses and $M_*=0.8M_\odot$. Bottom two panels: same results with $M_*=0.55M_\odot$. The simulations are performed using the parameters reported in \citet{rosenfeld_12}: $p=0.93$, $q=0.26$ and $H_{\rm in}/R_{\rm in}=0.1$. In the most extreme case, with the highest possible mass for the planet and the lowest possible mass for the star, the system would need a viscosity of $\alpha\approx0.15$ to explain the hypothesised warp in this framework. In this specific case the model requires the fractional amplitude of the warp to be greater than $0.1$, as shown by the horizontal black line. The dashed lines represent the degree of warping estimated using the analytic prescription from Eq. (\ref{eq:deltabeta}). For small warps ($\Delta\beta/\beta_\infty<0.1$) the agreement between simulations and analytic prescription is remarkable.}
\label{fig:tw_m080}
\end{figure*}

From \citet{rosenfeld_12} we know that the best agreement between the observations and the warped model indicates $H_{\rm in}/R_{\rm in}=0.1$, $p=0.93$ and $q=0.26$. We consider a disc with inner radius $R_{\rm in} = 4$ AU in gas, which appears to be the most plausible case. \citet{rosenfeld_12} used a central star with mass $M_*=0.8M_{\odot}$ in their modelling. However, there has been a long debate on the mass of the central star. We do not give the details here, we just report that mass estimates for the star from spectral diagnostics range between $0.5-0.8M_\odot$ \citep[e.g.][]{alencar_02,vacca_11}. \citet{debes_13} propose a mass estimate of $0.55\pm0.15M_\odot$ by using spectral signatures and the light scattered by the dust-laden protoplanetary disc. We have run simulations between the two extremes, by considering $M_*=0.8M_\odot$ and $0.55M_\odot$.
  
We estimate the semi-major axis of the planet's orbit by requiring $R_{\rm in}$ to be the tidal truncation radius and $M_*$ to be fixed in the centre of the planetary orbit. We can use this approximation since $M_*\gg M_{\rm p}$, where $M_{\rm p}$ is the mass of the planet. The Hill's radius of the planet, $R_{\rm H}$ is:
\begin{equation}
\label{eq:hill}
R_{\rm H} = \left(\frac{M_{\rm p}}{3M_*}\right)^{1/3}a\approx 0.18a\left(\frac{M_{\rm p}}{14M_{\rm J}}\right)^{1/3}\left(\frac{M_*}{0.8M_{\odot}}\right)^{-1/3},
\end{equation}
and the inner disc radius is related to the planetary semi-major axis by
\begin{equation}
\label{eq:tidal}
R_{\rm in} = a + R_{\rm H},
\end{equation}
We fix $R_{\rm in}=4$ AU, as observed. Thus, for any choice of $M_{\rm p}$ and $M_*$, we obtain $R_{\rm in}/a$. In this way, we have reduced our model to a two dimensional parameter space, where the tilt of the disc $\Delta\beta/\beta_\infty$ depends on $M_{\rm p}$ and $\alpha$ only, where the stellar mass has been fixed to either $0.55M_{\odot}$ or  $0.8M_{\odot}$. We recall that the steady-state solution of a circular circumbinary disc depends on $\chi$ and $\hat{\alpha}$, where $\chi$ is a function of both $\eta$ and $a$. Since we know $p$, $q$, $M_*$, $H_{\rm in}/R_{\rm in}$ and $a(M_{\rm p})$, the only two free variables are $M_{\rm p}$ and $\alpha$.

\citet{rosenfeld_12} obtain a good agreement in the line profiles and in the channel map for the $^{12}$CO $J=2-1$ and $^{12}$CO $J=3-2$ emission lines with an outer disc inclination $i_\infty \approx 4^\circ$ and an inner disc inclination $i_{\rm in}\approx 8^\circ$. Therefore, we model a disc with a warping factor of $\Delta\beta\approx 4^\circ$. These observations, however, do not fully determine the warp fractional amplitude, since we do not know what is the orientation of the planetary orbit with respect to the plane of the sky. Note that, since in this case $R_{\rm warp}<R_{\rm in}$, we do not expect the inner disc to be aligned with the planetary orbit. One additional unknown is thus the misalignment between the outer disc and the planetary orbit, $\beta_\infty$. However, SPH simulations by \citetalias{facchini13_1} have shown that if the misalignment of the outer disc is larger than $\approx40^\circ$, the evolution of the warp becomes non-linear. Even though the evolution of the system in the strongly non-linear case is not fully understood yet, \citetalias{facchini13_1} indicate that in these conditions the disc might break into two separate planes and we would thus not expect a smooth warp. Since \citet{rosenfeld_12} find a good agreement between the observed channel maps and a warped model with an inclination of the disc being a smooth function of $R$, we conservatively require that $\beta_\infty\lesssim 40^\circ$ (i.e. in the linear regime), so that $\Delta\beta/\beta_\infty\gtrsim 0.1$.

In Fig. \ref{fig:tw_m080} we show $\Delta\beta/\beta_\infty$ as a function of $\alpha$ for the two estimates of the star mass ($M_*=0.8M_\odot$ and $0.55M_\odot$, respectively), where different lines depict different planet masses (from $1M_{\rm J}$ to $20M_{\rm J}$). The horizontal black line indicates the threshold $\Delta\beta/\beta_\infty=0.1$. The  green line shows the $M_{\rm p} = 14 M_{\rm J}$ case, which is the observational upper limit for the mass of the planet. The dashed lines represent the analytic prescription given the used disc's parameters. The agreement between the analytic fitted prescription and the simulations for $\Delta\beta/\beta_\infty< 0.1$ is good. Note that these results do not depend on the estimate of $R_{\rm in}$. The amount of warping depends on $a/R_{\rm in}$, which is set by the tidal truncation radius condition (it is not a function of $R_{\rm in}$ itself, see Eqs. (\ref{eq:tidal})-(\ref{eq:hill})).

\begin{figure*}
\begin{center}
\includegraphics[width=.95\columnwidth]{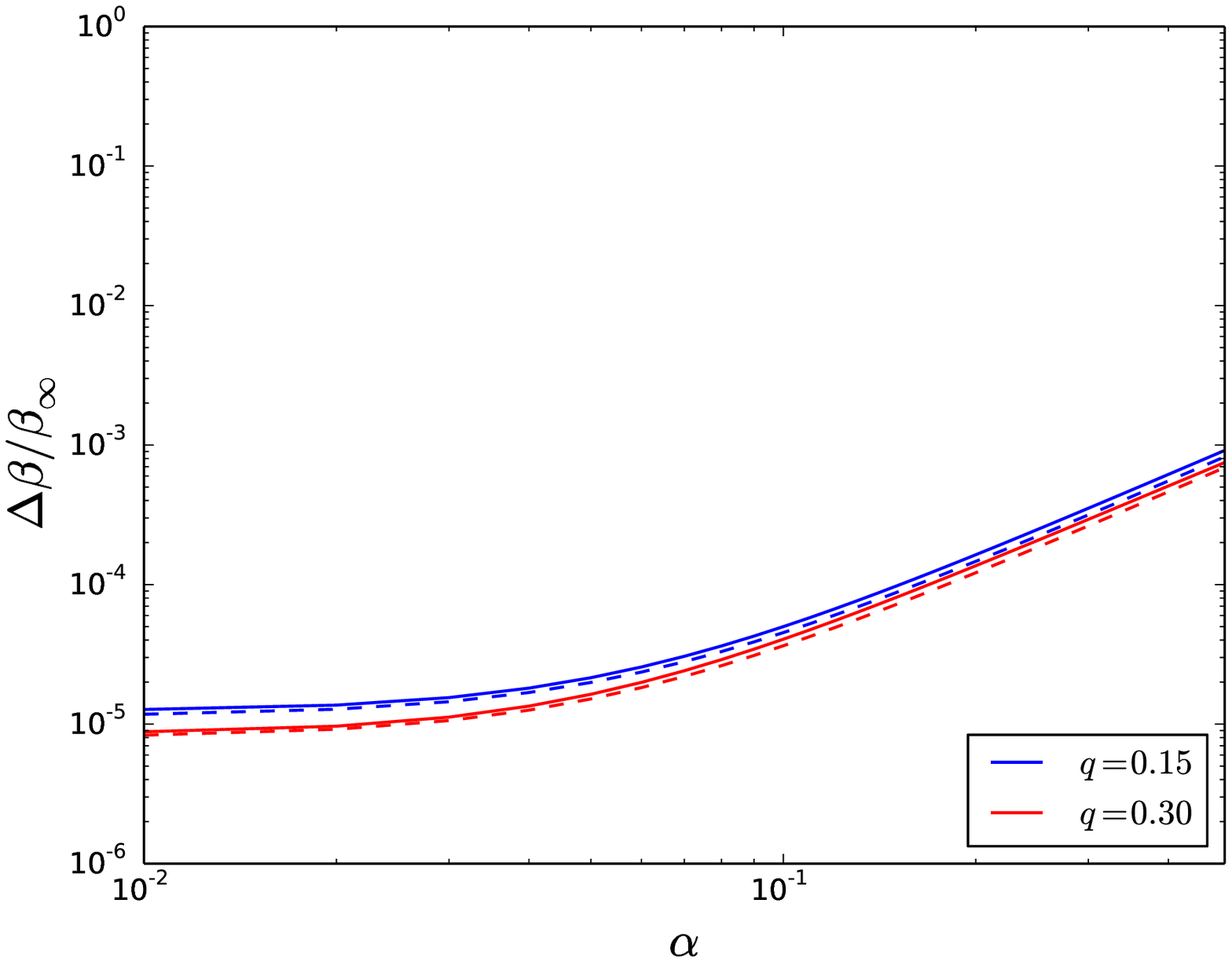}
\includegraphics[width=.95\columnwidth]{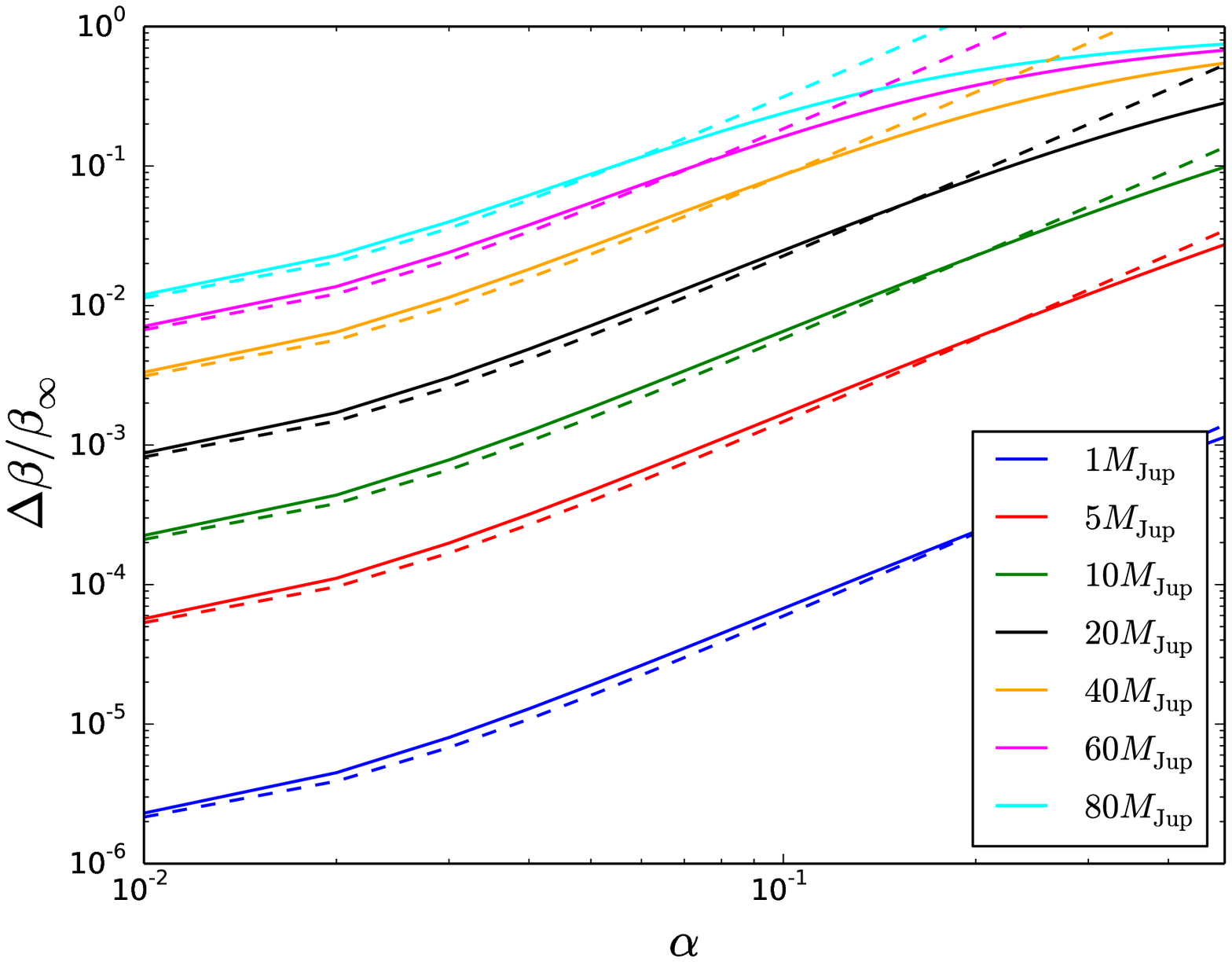}\\
\includegraphics[width=.95\columnwidth]{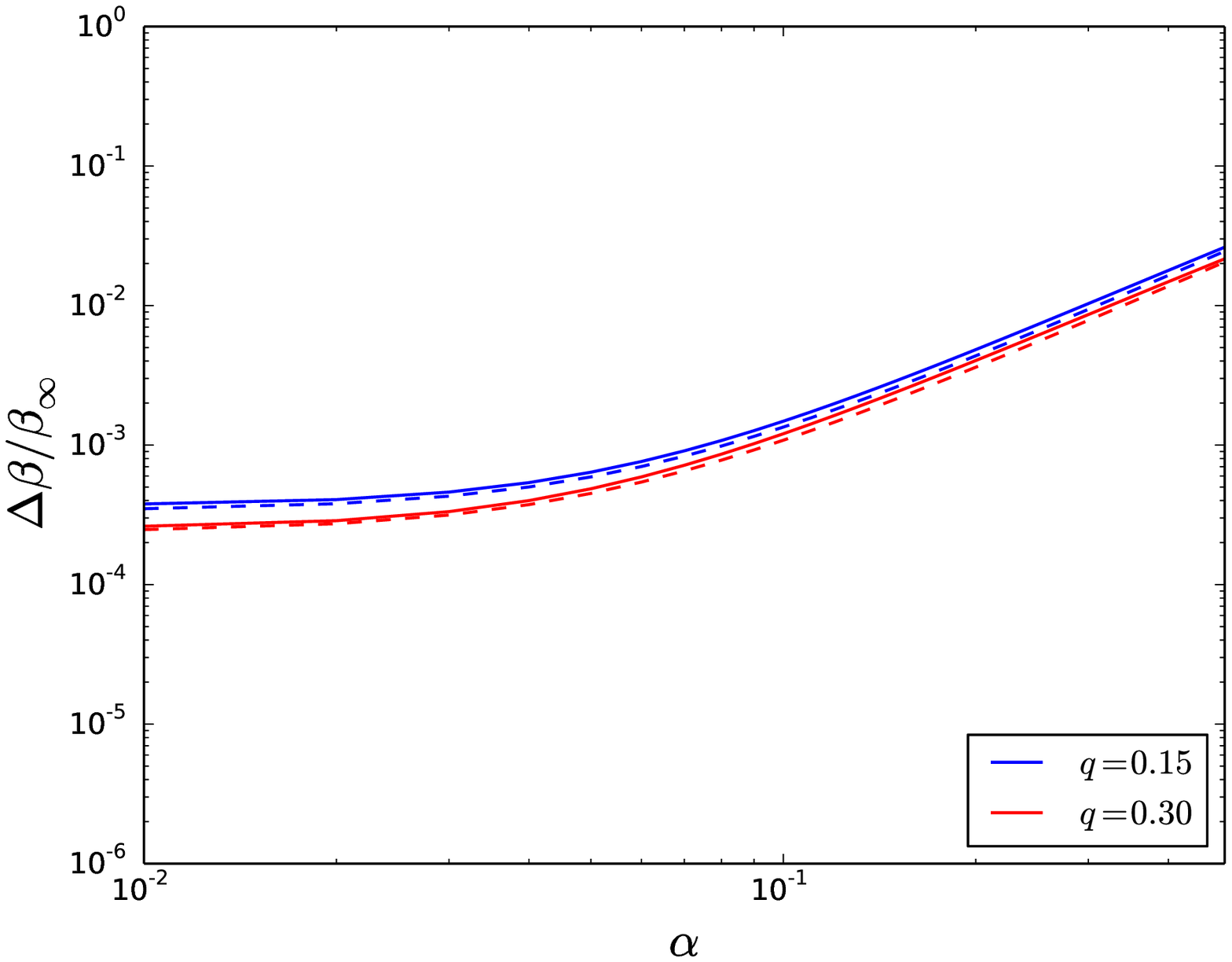}
\includegraphics[width=.95\columnwidth]{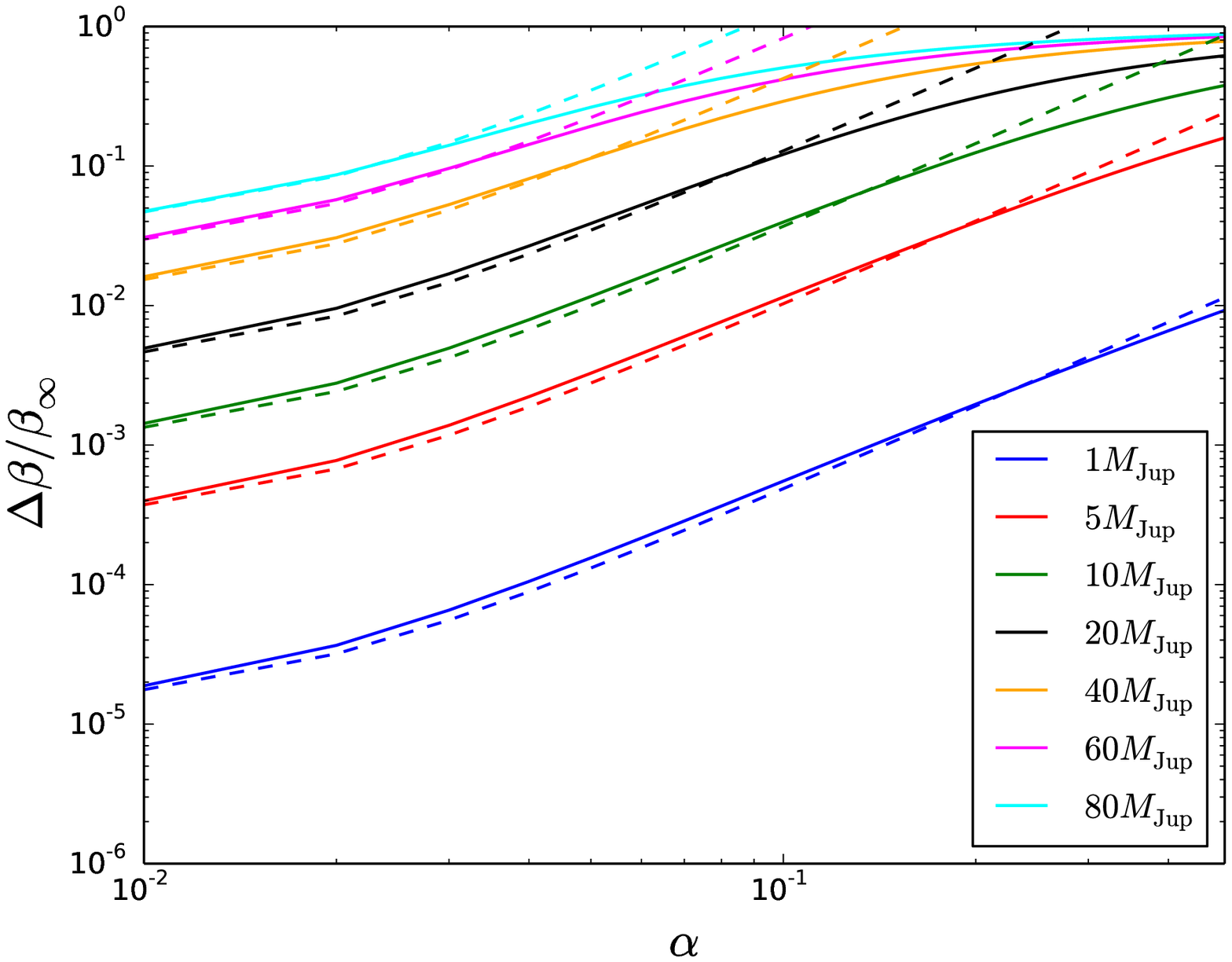}
\end{center}
\caption{Top left: $\Delta\beta/\beta_\infty$ as a function of viscosity for different values of $q$ in the case of LkCa 15. Top right: $\Delta\beta/\beta_\infty$ as a function of viscosity for different values of $M_{\rm p}$ in the case of T Cha. The two bottom panels show the results of simulations equivalent to the ones of the top panels, but where $R_{\rm in}$ is estimated via Eq. (\ref{eq:tidal}), given the observed value of $a$. For LkCa 15 the expected warp is negligible in both cases. In the case of T Cha, if the companion has a mass of at least $20M_{\rm J}$ and sits on an inclined orbit with respect to the disc, an observable warp would be expected for reasonable values of the viscosity. As in Fig. \ref{fig:tw_m080}, the dashed lines represent the degree of warping estimated using the analytic prescription from Eq. (\ref{eq:deltabeta}). For small warps ($\Delta\beta/\beta_\infty<0.1$) the agreement between simulations and analytic prescription is good again.}
\label{fig:lkca15}
\end{figure*}

Both plots at different stellar masses show that in order to reach the minimum value of $\Delta\beta/\beta_\infty\approx 0.1$, as discussed above, the planet needs to be relatively massive, at least $5M_{\rm J}$, in the range of explored viscosities. Additionally, it is also apparent that the disc viscosity needs to be high.  Even when $M_*=0.55M_\odot$ and $M_{\rm p}=14M_{\rm J}$, $\Delta\beta/\beta_\infty\gtrsim 0.1$ for $\alpha\gtrsim 0.15$. Therefore in the most extreme case, with the highest possible mass for the planet and the lowest possible mass for the star, the system would still need a viscosity of $\alpha\approx0.15$ to explain the observed warp with this model. However, note that $\Delta\beta/\beta_\infty$ scales approximately with the inverse fourth power of $H_{\rm in}/R_{\rm in}$ (Eq. (\ref{eq:deltabeta})). Thus, even a slightly smaller value for the disc aspect ratio would require a much smaller value of $\alpha$ in order to reach the same degree of warping.

As mentioned in Section \ref{sec:theory}, the viscosity of protoplanetary discs is expected to be in a range between $10^{-4}-10^{-2}$ \citep{hartmann_98}, significantly below our estimate. For those systems were $\alpha$ can be estimated more accurately \citep[i.e., for X-ray binaries][]{king_07}, the resulting values are in line with our inference, but the gas in these systems is fully ionised so that the magneto-rotational instability can fully operate. For protoplanetary discs the ionisation level is expected to be low, which would result in a much lower effective viscosity. Still, additional transport processes might play a role in cold protostellar discs, such as gravitational instabilities \citep{lodato_04,cossins_09} that are able to provide effective viscosities of the order of $\alpha\approx 0.1$ \citep{rice_05}. In order for gravitational instabilities to be effective, the ratio of the disc to stellar mass needs to be of the order of the aspect ratio. For TW Hya, \citet{debes_13} estimate a disc mass of $M_{\rm d}\approx 0.027M_{\odot}$ between 27 and 211 AU, and \citet{bergin_13} give a even higher mass $M_{\rm d}\gtrsim 0.05M_{\odot}$, that implies $M_{\rm d}/M_*\approx 0.1$. Thus we do expect the disc to be marginally gravitationally unstable. It should be noted however that, as it is often the case, such disc mass estimates are very uncertain and might be affected by strong systematic errors \citep[e.g.][recently estimated a disc mass of $\sim6\cdot10^{-4}M_\odot$ by invoking a very high dust-to-gas ratio]{williams_14}. Finally, the very high $\alpha$ value required by our model seems to be in contradiction with such a massive disc at late stage in its evolution \citep[$\sim10$ Myr,][]{barrado_06}. This issue has been widely discussed in the literature, and there is no clear answer yet. We do not want to discuss this problem in detail here. Note however that the age estimates are usually derived from the whole TW Hydrae Association. \citet{weinberger_13} has recently shown that the stars in the association might not be coeval, therefore implying large uncertainties in the age estimates of the single stars. \citet{debes_13} estimated the age of the star TW Hya by modelling the stellar spectrum, and they find that the best interpretation of the data is given by an age of $8\pm4$ Myr. Therefore the friction between age and high viscosity could be attenuated if TW Hya lay in the young tail of the distribution.

In our model, the planet is simultaneously inducing the warp and tidally clearing out the central cavity. Since we requested $\alpha$ to be very large in order to model the observations by \citet{rosenfeld_12}, we need to check whether the inferred planet would still be able to open up a gap into the disc. In order a gap to be opened, the mass ratio between the planet and the star has to be larger than a critical value $q_{\rm crit}$, expressed by a simple combination of pressure and viscous terms \citep{armitage_book_10}:

\begin{equation}
\label{eq:q_crit}
\frac{M_{\rm p}}{M_*}>q_{\rm crit}=\left(\frac{27\pi}{8}\right)^{1/2} \left(\frac{H}{R}\right)^{5/2}\alpha^{1/2}.
\end{equation}
This criterion is verified by the values we use in our model ($\alpha=0.15$ and $H/R=0.1$) for all planet masses, except for the case $M_{\rm p}=1M_{\rm J}$. By assuming $M_*=0.55M_\odot$, we can express Eq. (\ref{eq:q_crit}) as $M_{\rm p}/M_{\rm J}>2.29$. Even with such large values of viscosity, the planet is still able to clear out a gap, since the planet itself needs to be very massive in order to induce a detectable warp.

Another critical point is whether the planet is able to maintain its inclination for long timescales. \citet{terquem_13} has recently shown that planets that orbit within a central cavity can maintain a high inclination with respect to the disc for timescales longer than the disc lifetime ($>10$ Myr). Since the planet is well inside the cavity, there is no frictional force that damps the orbit towards the disc plane \citep[e.g.][]{teyssandier_13,xiang_13}. Therefore, if the planet reaches an inclined orbit within a central cavity of a disc via gravitational interaction or via secular perturbations, it can maintain such an inclination long enough to be statistically observable.

\begin{table}
\centering
\begin{tabular}{lccc}
\hline
		                     		& TW Hya 	& LkCa 15 	& T Cha		 \\
\hline
$M_*$ ($M_\odot$)			&	0.55-0.8	&	1.0		&	1.5		\\
$p$						&	0.93		&	0.72		&	1		\\
$q$						&	0.26		&	-		&	0.4		\\
$H_{\rm in}/R_{\rm in}$		&	0.1		&	0.18$\cdot$(0.65)$^{-q}$		&	0.078	\\
$M_{\rm p}$ ($M_{\rm J}$)	&	$<$14	&	6		&	$<$80	\\
$a$ (AU)					&	-		&	16		&	6.7		\\
$R_{\rm in}$ (AU)			& 	4		&	42		&	12		\\
$a+R_{\rm H}$ (AU)			& 	-		&	18.0		&	7.1-8.4		\\
References				&	1,2,3,4	&	5,6,7,8	&	9,10,11	\\
\hline
\end{tabular}
\caption{Parameters used in the simulations shown in Figs. \ref{fig:tw_m080}-\ref{fig:lkca15} for TW Hya, LkCa 15 and T Cha, respectively. We list all the parameters that are constrained by observations, expect for $a+R_{\rm H}$, which is evaluated from the mass and the orbital radius of the substellar companion. To estimate $a+R_{\rm H}$ for T Cha, where only an upper limit on the mass of the companion is available, we have used masses between $1M_{\rm J}-80M_{\rm J}$. \emph{References}: (1) \citet{rosenfeld_12}; (2) \citet{andrews_12}; (3) \citet{evans_12}; (4) \citet{debes_13}; (5) \citet{pietu_07}; (6) \citet{andrews_11}; (7) \citet{isella_12}; (8) \citet{kraus_12}; (9) \citet{huelamo_11}; (10) \citet{cieza_11}; (11) \citet{oloffson_13}.}
\label{tab:discs}
\end{table}

\subsection{LkCa 15}
\label{sec:lkca15}

LkCa 15 is a transition disc in the Taurus-Auriga star forming region \citep[at a distance of $\sim$ 145 pc,][]{torres_09} with an inner cavity of roughly 45 AU \citep{espaillat_07,andrews_11,isella_12}. Recently, \citet{kraus_12} have observed through aperture masking interferometry a protoplanet located at a distance of $\sim 16$ AU from the star (for a circular orbit coplanar with the disc), that is, well within the cavity. The mass estimate for the planet is $6M_{\rm J}$, with an upper limit of 12$M_{\rm J}$. The stellar mass is $\sim 1M_{\odot}$ \citep{simon_00}.  No warp has been observed in the disc. The inclination of the disc is $49^{\circ}$ \citep{andrews_11}. 

Here, we estimate the expected $\Delta\beta/\beta_\infty$ that would occur (for a linear warp) if the planetary orbit were inclined with respect to the disc. The assumed parameters for the disc are summarised in Table \ref{tab:discs}. The resulting fractional warp amplitudes are shown in Fig. \ref{fig:lkca15} (top left panel) as a function of $\alpha$ for the extreme values of the power-law index for the sound speed $q$ in the range predicted by \citet{pietu_07} ($q=0.15$ and $q=0.30$). For any reasonable choice of parameters the expected warp amplitude is extremely low ($\lesssim 10^{-3}$) and we thus do not expect to observe a warp even if the misalignment of the orbit is large. Note however that the inner edge of the disc is much larger than the expected tidal radius of the planet. This could be due to the fact that the estimates of the inner radius of the disc come from millimetric observations, therefore tracking dust grains with a size of $\approx 1$ mm. Gas (and small dust) could extend well within the millimetric dust cavity \citep[e.g.][]{pinilla_12}, as observed in some systems \citep[e.g.][]{garufi_13}, where the gaseous inner radius is estimated via SED modelling in the IR or via scattered light measurements (therefore tracking micron-size dust, which is dynamically well coupled with the gas). We have therefore re-run the simulations by assuming a gaseous inner radius given by Eq. (\ref{eq:tidal}). The results are shown in the bottom left panel of Fig. \ref{fig:lkca15}. The expected fractional warp increases by a factor of $\sim15$. However, for this system, it remains under the detectability threshold.

\subsection{T Cha}
\label{sec:t_cha}

T Cha is a transition disc at a distance of $\sim$ 108 pc \citep{torres_08} with an inner cavity of $\sim$ 12 AU \citep{oloffson_13}. A companion candidate has been observed at a separation of $6.7$ AU \citep{huelamo_11}. In this case there is no mass estimate for the companion.  The stellar mass is $\sim 1.5M_{\odot}$ \citep{alcala_93}.  Also in this case, no warp has been observed in the disc. The inclination of the disc is $60^{\circ}$ \citep[e.g.][]{cieza_11}. The assumed parameters for the disc are summarised in Table \ref{tab:discs}. 

The resulting fractional warp amplitudes are shown in Fig. \ref{fig:lkca15} (right panels) as a function of $\alpha$ for different values of the companion mass ranging from a gaseous giant up to a brown dwarf mass. The top panel shows simulations where $R_{\rm in}$ is equal to the observed value, whereas in the bottom panel $R_{\rm in}$ is estimated with Eq. (\ref{eq:tidal}). In this second case a sizable warp is expected if the companion mass is larger than 40$M_{\rm J}$ for $\alpha\sim 0.06$. For a less viscous disc, the warp amplitude is expected to be lower and thus only companions with high mass (in the brown dwarf regime) can be expected to produce an observable warp.

\section{Conclusions}
\label{sec:disc}

In this paper we have shown how warped circumbinary discs can be used to study some key physical properties of protoplanetary discs. We have reported a simple prescription that can be used to relate the amount of warping of these discs with the dynamical properties of the two central objects. This relation generalises previous results \citep{foucart_lai_13} to a wider parameter range and to cases where either the binary torque or the disc viscosity are large enough to lead to the alignment of the inner disc, a situation that has been neglected before. These results are general, and they can be applied both to stellar binaries, to a star and a brown dwarf, and to a central star and a planet. 

Observing a misaligned protostellar circumbinary disc would give new constraints on the magnitude of the disc's viscosity \citep{king13}. As reported in Section \ref{sec:introduction}, existing facilities have the capability to measure warps in protostellar discs by looking at the kinematics of the gas. Once the systemÕs dynamical parameters are known (for example, the masses and orbit of the central binary) the other parameter that determines the warp amplitude is the disc viscosity, so that by measuring the amount of warping in a disc it is possible to have an indirect measure of the disc's viscosity itself. Note that a similar approach has been recently attempted \citep[e.g.][]{king13,facchini13_2}, by studying the alignment timescale for either a circumbinary or a circumprimary disc.

We then applied such 1D model to TW Hya. In this system, \citet{rosenfeld_12} have measured a velocity pattern that is not consistent with a flat Keplerian disc and have hypothesised that a warp is present in the disc, whose inclination changes by 4 degrees between the inner and outer radius. TW Hya is also a transition disc, with a large inner hole that can be caused by the tidal interaction with a massive planet $M_{\rm p}\lesssim 14 M_{\rm J}$. Here, we investigate the possibility that the same planet is also responsible for the warp in the disc. A major piece of information that is missing from observations is the orientation of the orbit of the purported planet. Here, we have assumed that the warp is linear and thus that the misalignment of the outer disc cannot be too large (we conservatively require $\beta_\infty\lesssim 40^{\circ}$), which then implies that $\Delta\beta/\beta_\infty\gtrsim 0.1$. Note that in the linear assumption, the warped structure is smooth in its radial dependence, as required by \citet{rosenfeld_12} to obtain a good agreement between the observations and a simple warped disc model. We then investigated what disc and orbital parameters are required to reproduce such degree of warping. We conclude that the planet needs to be relatively massive and the disc needs to be relatively viscous. For $M_{\rm p}=10-14M_{\rm J}$, we require $\alpha\approx 0.15-0.25$. These results depend sensitively on the assumed aspect ratio of the disc ($H_{\rm in}/R_{\rm in}=0.1$) and our constraints would be lessened if the disc was thinner. Such a large viscosity is not expected to be produced by the magneto-rotational instability in cold and weakly ionised protostellar discs, but can potentially be explained if the origin of viscosity is related to gravitational instabilities.

We expect that with the new ALMA observations, this kind of warped structures in protoplanetary discs will be observed more frequently and we thus expect our method to be applied to a much wider sample in the future. This method will favour systems with some characteristic features. Using the gas emission lines profile to measure the warp, as described by \citet{rosenfeld_12}, will favour closely face-on discs. In this condition $\sin{i}\sim i$, therefore the projected radial velocity will be linearly sensitive to the inclination angle. Secondly, independently from the used technique, at a given mass ratio between the central star and the companion, the warping will be more pronounced for low values of the scale-height  $H$, i.e. for thinner and colder discs. Finally, the warp will be more prominent for a larger mass ratio (where the mass ratio is defined as $M_p/M_*$). Therefore, for a  given planet mass, low mass stars (as M dwarfs) will be favoured.

\section*{Acknowledgements}
We thank the referee, whose comments helped improving the manuscript. We thank Cathie Clarke for stimulating discussion. SF thanks the Science and Technology Facility Council and the Isaac Newton Trust for the award of a studentship. GL acknowledges financial support from PRIN MIUR 2010-2011, project ``The Chemical and Dynamical Evolution of the Milky Way and Local Group Galaxies'', prot. 2010LY5N2T.

\bibliography{tw_hya_bib}

\label{lastpage}
\end{document}